# Model-based testing in practice: An experience report from the web applications domain

| Vahid Garousi<br>Queen's University Belfast, UK<br>Bahar Software Engineering Consulting Corporation, UK<br>v.garousi@qub.ac.uk | Alper Buğra Keleş, Yunus Balaman, Zeynep Özdemir Güler<br>Testinium A.Ş., Istanbul, Turkey<br>{alper.keles, yunus.balaman, zeynep.ozdemir}@ testinium.com | Andrea Arcuri<br>Kristiania University College, Norway<br>Oslo Metropolitan University, Norway<br>andrea.arcuri@kristiania.no |
|---|---|---|

**Abstract:** In the context of a software testing company, we have deployed the model-based testing (MBT) approach to take the company's test automation practices to higher levels of maturity and capability. We have chosen, from a set of open-source/commercial MBT tools, an open-source tool named *GraphWalker*, and have pragmatically used MBT for end-to-end test automation of several large web and mobile applications under test. The MBT approach has provided, so far in our project, various tangible and intangible benefits in terms of improved test coverage (number of paths tested), improved test-design practices, and also improved real-fault detection effectiveness. The goal of this experience report (applied research report), done based on "action research", is to share our experience of applying and evaluating MBT as a software technology (technique and tool) in a real industrial setting. We aim at contributing to the body of empirical evidence in industrial application of MBT by sharing our industry-academia project on applying MBT in practice, the insights that we have gained, and the challenges and questions that we have faced and tackled so far. We discuss an overview of the industrial setting, provide motivation, explain the events leading to the outcomes, discuss the challenges faced, summarize the outcomes, and conclude with lessons learned, take-away messages, and practical advices based on the described experience. By learning from the best practices in this paper, other test engineers could conduct more mature MBT in their test projects.

**TABLE OF CONTENTS**

## 1 INTRODUCTION

Systematic and adequate testing of software systems is a costly activity, but so do the costs caused by software defects due to inadequate testing. In a quest to increase effectiveness and efficiency of testing, software engineers have used test automation [1] for several decades now. While most practitioners use automation for the test execution phase, test automation is "*not just for test execution*" [2], i.e., it can be used in other test activities such as test-case design.

Model-based testing (MBT) [3] is an established black-box testing approach for generation of test cases. In MBT, specific types of models, often called *test models*, are developed or are reused from earlier software lifecycle phases (e.g., requirements or design) for generation of test cases. When MBT is integrated with test execution tools such as Selenium for web applications, it can also automate execution of test cases derived from test models, thus further increasing effectiveness and efficiency of testing.

MBT has been around for at least 50 years now. An IBM technical report [4], published in 1970, is often referred to as one of the first known reported applications of MBT. The modeling semantic (type of test models) followed in that first paper was Cause-Effect Graphs, and a prototype tool, named TELDAP (TEst Library Design Automation Program), for generating test cases was presented. A very large number of papers and reports have been published in MBT since then, by following different approaches to MBT, e.g., from the standpoints of model semantics (UML models, BPMN or other model types), level of modeling abstractions, test execution modes (offline or online), and test selection criteria (model coverage, fault-based, etc.) [5-7]. However, many studies report that: "*most developers [still] don't view MBT as a mainstream [testing] approach*" [8].

Specific domains have historically used and taken more advantage of MBT, e.g., embedded software, aerospace, railway and telecommunications [5]. While test teams in the above specific domains often have the resources to adopt/build domain-purpose (and often heavy-weight) MBT approaches, adopting MBT in the enterprise software domains, e.g., web and mobile applications, has not been successful with heavy-weight approaches and, instead, needs lean, highly usable, lightweight and cost-effective methods and tools [9].

In the context of a software testing company (Testinium A.Ş.) with offices in several European countries, we have pragmatically used MBT, since January 2019, to improve the company's test-automation practices. The work is the result of an industry-academia collaboration [10], and has been conducted in the context and using the funding of an international large European R&D project named "TESTOMAT – The Next Level of Test Automation" (testomatproject.eu), in which 34 industrial / academic partners across six countries are collaborating. The TESTOMAT project ran from 2018 to the end of 2020. To provide the larger context of the work reported in this paper, let us note that MBT is only one of the work-packages of the TESTOMAT project, and in the industrial context of the subject company (Testinium A.Ş.), several other test automation innovation have also been conducted and published as recent papers, e.g., experience reports and a set of innovative best practices for executable natural-language test specifications using a test tool called *Gauge* (gauge.org) were published in [11, 12].

Given the very large number of MBT approaches and tools [6, 7], our goal in the TESTOMAT project has been not to develop a yet new MBT approach, but rather to select and apply the "right" MBT approach(es) in the context of the subject company (Testinium A.Ş.), to identify the practical challenges/questions that a typical company or test team would face when deploying MBT in practice in the context of web and mobile applications, and to take the company's test automation practices using MBT to higher levels of maturity and capability.

In this paper, we report on the experience of a project on choosing and applying a practical MBT approach in practice, the insights that we have gained, and the questions and challenges that we have faced so far, e.g., which MBT approach/tool should we choose? How to deploy a lightweight MBT approach in our context?

Since this is mainly an industrial project and had to deliver improvements in practice, our approach has been "pragmatic". In discussion with company's management, from the beginning of the project, it was clear that we could not use "heavyweight" MBT and Model-Driven Engineering (MDE) approaches that would require extensive modelling without considering their cost-benefits in practice [13]. For example, we had to ensure that the chosen modeling is as simple as possible, to ensure ease of adaption in test teams. Our project's philosophy has been similar to that of another experience report on applying MBT in industry [13], in which the author argued that "*it is important to always state where the models [to be used in MBT] come from: are they artificial or did they already exist before the experiments*" and that "*one has to argue and evaluate if the time and effort in developing and maintaining such models for a given system does pay off in the end*".

The remainder of this paper is structured as follows. Since we used Gorschek et al.'s process model [14] in our project (details in Section 3), sections of this paper are structured based on that process. Section 2 reviews the industrial context, needs and the motivations for the project. We discuss the research approach, design and questions of the project in Section 3. In Section 4, we review the related work. As the core of our work, our test automation strategy and test-artifact development are discussed in Section 5. We report in Section 6 the empirical findings that we have gathered so far in the project, for assessing the (positive) impacts and the benefits that MBT had in our project, and also the challenges and questions that we have observed so far. In Section 7, we discuss the lessons learned, take-away messages, and practical advice based on the described experience. Finally, Section 8 concludes the paper and discusses our current and future work directions.

## 2 INDUSTRIAL CONTEXT, NEEDS AND MOTIVATIONS FOR THE PROJECT

Testinium A.Ş. is officially classified as a Small/Medium-sized Enterprise (SMEs). The company employs more than 200 software test engineers. Almost all test engineers have received different certificates types of the ISTQB (International Software Testing Qualifications Board), e.g., the "Foundation Level" certificate.

The company has been proactive in adapting novel approaches to increase effectiveness and efficiency of its test activities, and joining the European TESTOMAT project has been one of those initiatives. Almost all of the Systems Under Test (SUTs) tested by test engineers are the clients' web or mobile applications, e.g., the online ticket sales website of several major airlines in Turkey.

Two major system GUI-level automated testing technologies used in the company are *Selenium* (selenium.dev) and *Gauge* (gauge.org). System GUI-level testing is to conduct system testing on a SUT via its Graphical User Interface (GUI). While such tools are effective for automated execution of the developed test scripts, based on our many test automation projects, we and many others [15] have found those test tools *alone* are not *enough* for a successful test automation outcome. A critical issue is that the automated test artifacts should be designed and developed properly, e.g., should be free from "test smells" [16] and should be modular, since as test code grows, it becomes a code-base of its own. Furthermore, the test cases underlying the test scripts should be systematically designed to have the most cost-effective test suites, i.e., the most-size-optimal test suites having the highest fault detection effectiveness. Furthermore, test scripts have to be maintainable, since the requirements, code-base and/or the GUI of the SUT often change. Doing all these aspects in a disciplined manner was referred to as "software test-code engineering" (STCE) in our previous work [17]. An industry expert, named Hans Buwalda, also summarizes this point clearly as: *"Success in automation is not as much a technical challenge as it is a test design challenge"* (bit.ly/TestDesignForAutomation).

In our industrial context (Testinium A.Ş.), various black-box test design approaches have been in use since the company was founded in 2010, e.g., category-partition testing and boundary-value testing. However, since such techniques can be interpreted and applied in different ways by different test engineers, the automated test suites were designed in different ways and we have been seeing the need for a "better" test-design approach. Furthermore, although there has been a very large research literature on test-design in academia (like those discussed above, e.g., category-partition testing), systematic test-case design practices do not seem to be in wide use in many industrial contexts [18]. This has mainly been attributed to low applicability of textbook-based test-design approaches in practice [18].

Based on the above exploratory phase and needs analysis in Testinium A.Ş., and by reviewing the experience report and success stories of MBT in practice, e.g., [8], we selected MBT to improve the test-case design and test automation practices, which was also raised as one of the work-packages of the TESTOMAT project (testomatproject.eu).

## 3 PROJECT PROCESS AND ACTION-RESEARCH QUESTIONS

In terms of research process for the project and our industry-academia collaboration [10], we used the widely-cited process model proposed by Gorschek et al. for action-research and technology transfer in SE [14], which consists of seven steps: (1) Identify the industrial need(s), through assessment and observation activities; (2) Formulate a research agenda by reviewing the state-of-the-art (literature) and -practice to find the research focus; (3) Formulate a candidate solution in cooperation with industry; (4) Conduct lab validation (for example, through lab experiments); (5) Perform static validation in the industrial context (for example, via interviews and seminars); (6) Perform dynamic validation (for example, pilot projects); and (7) Release the solution in the industrial context step by step, while remaining open to smaller changes and improvements. That process model [14] has been used widely cited in the literature and has been used in a large number of industry-academia collaborations, e.g., our past projects with a large number of partners, e.g., [10, 19]. For our research process, we also benefitted from other papers and guidelines for action research, e.g., [20-23].

Our project goal was to assess practical applicability and cost-effectiveness of MBT in the industrial context by applying it to several large testing projects, with the hope of making MBT a common test-automation approach in the company. We believe that sharing our success story would motivate practitioners for using MBT.

Given the very large spectrum of MBT approaches and tools [6, 7], we had to choose and adapt the right MBT approach and tool, by taking advice from an insightful voice-of-evidence paper [8] which mentioned: "*Developers must obviously take care to select an MBT approach that matches their project's specific needs*".

Furthermore, using any software engineering (SE) approach in practice by any SE team has non-trivial costs, and the associated cost-benefits should be carefully analyzed, a topic referred to as "value-based" software engineering [24]. Only if benefits of a given SE approach outweigh its costs, a given SE team will decide or continue using it. A paper by Neto et al. [8] confirmed this issue by stating that: "*it's risky to choose an MBT approach without having a clear view about its complexity, cost, effort, and skill required to create [develop] the necessary models*" and that: "*Evidence on these topics could be a useful step in determining whether wider deployment of MBT approaches to different domains is worthwhile*". We aimed at assessing these issues and to contribute evidence to the state of practice in this area, since studies have reported "*a serious lack in evidence*" [25] in MBT.

In the planning phase of our project, we derived the following three Action-Research Questions (ARQ), and we will address them in this paper:

- ARQ1: How can we choose the "right" MBT test tool for our purpose? (discussed in Section 5.2)
- ARQ2: What benefits does the MBT approach provide in the industrial context? (discussed in Section 6.1)
- ARQ3: Which challenges and questions did we face in the MBT project (so far) and how can they be addressed? (discussed in Section 6.2)

## 4 BACKGROUND AND RELATED WORK

By a literature search, one can find out that, since the first known MBT paper, published in 1970 as an IBM technical report, a few thousand papers have been published in various topics of MBT. Several survey and systematic review papers have summarized such a large body of knowledge, e.g., [6, 7, 25].

In the very large research literature and many books on MBT, we found that various MBT books and papers differ in terms of how applied and practical they are. We found the book by Kramer and Legeard [26] especially useful during our work, since it provides concrete, practical and pragmatic experience-based heuristics and guidelines for MBT.

In the rest of this section, we present:

- An overview of how MBT works
- State of the -art and -practice of MBT tools in general, tools for web applications, and types of test models
- MBT literature in practice and industrial contexts (since our work falls in this category)
- MBT body of knowledge in the Formal Methods community

### 4.1 An overview of how MBT works

Model-based testing (MBT) [3] is an established black-box testing approach for generation of test cases. In MBT, specific types of models, often called *test models*, are developed or are reused from earlier software lifecycle phases (e.g., requirements or design) for generation of test cases. When MBT is integrated with test execution tools such as Selenium for web applications, it can also automate execution of test cases derived from test models, thus further increasing effectiveness and efficiency of testing.

A UML activity diagram showing the general context and general process of MBT (taken from [27]) is shown in Figure 1. As discussed above, specific types of software models, often called *test models*, e.g., UML state-charts, are developed or are reused from earlier software lifecycle phases, e.g., requirements or design (forward engineering). There have been also many studies which have offered approaches for reverse engineering of (inferring) MBT models from code or other software artifacts, e.g., [28-30]. Those test models specify the expected behavior of the SUT. Once test models are ready and have been verified and validated, they can be used to derive test cases, which can then be executed on the SUT.

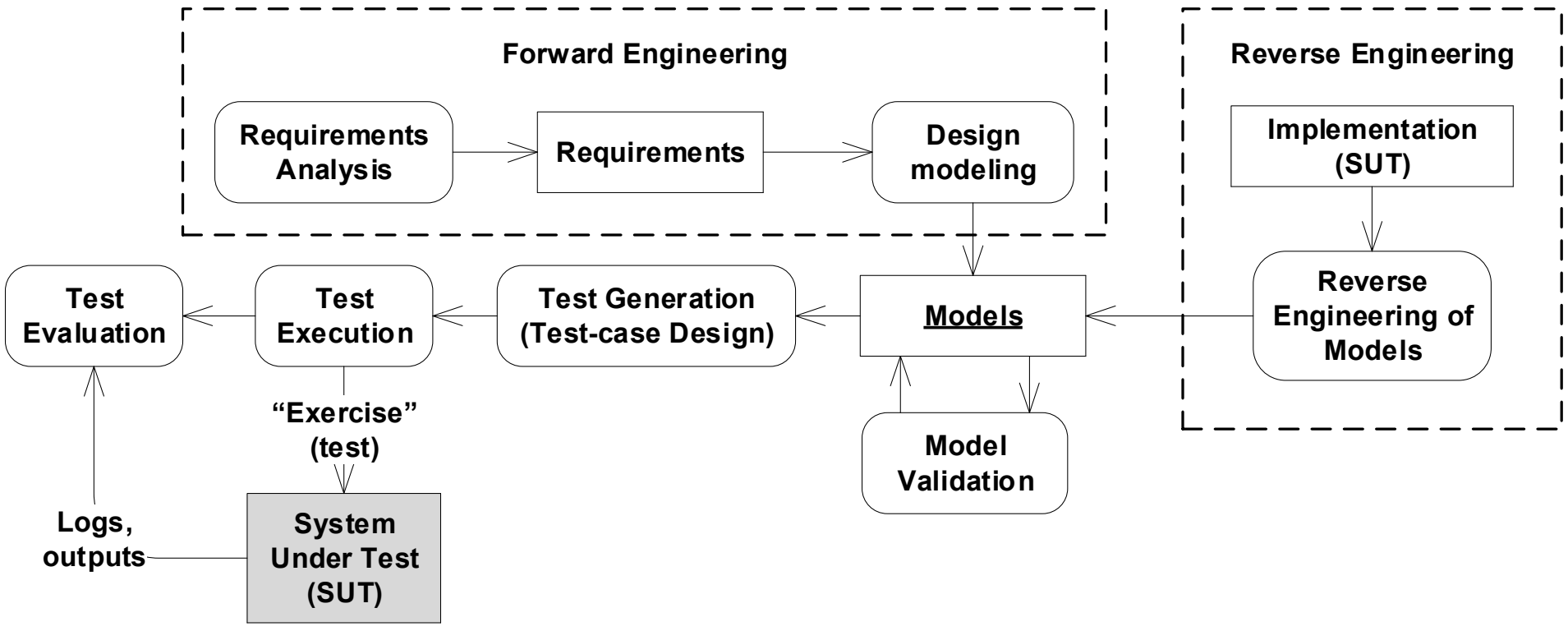

**Figure 1- A UML activity diagram showing the general context and general process of MBT [27]**

## 4.2 State of the -art and -practice of MBT tools in general, tools for web applications, and types of test models

There are perhaps hundreds of MBT tools, each specific to a certain domain and types of SUT's, e.g., mobile and web application, automotive software, etc. Even, surveys and systematic review on MBT tools and comparing their features have been published, e.g., [6, 7, 31]. MBT tools are often classified and compared by their supported type(s) of test models, test-generation criteria, and their test scripting capabilities [32]. Classification of several example (randomly-chosen) MBT tools, as presented in [32], is shown in Table 1.

**Table 1- Classification of several example (randomly-chosen) MBT tools, as presented in [32]**

| Tool name | URL | Target domain | Type of test model | Test-generation criteria | Test scripting capabilities |
|---|---|---|---|---|---|
| Conformiq Creator | www.conformiq.com | Web, desktop applications or web services | State charts | Requirements-driven test generation, black-box test design heuristics | Textual test plans and executable test cases in Java, and so on |
| Spec Explorer 2010 | https://research.microsoft.com/en-us/projects/specexplorer | Generic (applicable to all software domains) | State charts (Spec#) | Transition coverage | Executable test cases in C# or on-the-fly testing |
| MaTeLo | www.all4tec.net | Embedded software | Enhanced Markov chains | Probabilities for transitions and inputs | Textual test plans and executable test cases in TTCN-3 |

A subset of MBT tools are applicable to web applications. Given the nature of web applications, they are event-based systems, e.g., any mouse click on a hyperlink or HTML button in a given web page will change the page, and also the "state" of the web app under test.

By reviewing "survey" papers in this area [6, 7, 31] and also some exploratory Google searching, one can find a large list of MBT tools which can be used to test web applications. The following list of tools is a partial randomly-chosen subset:

- Commercial tools: TestModeller (testmodeller.io), TestOptimal (testoptimal.com), Tricentis Tosca (tricentis.com), etc.
- Open-source/free tools, made in industry: fMBT (github.com/intel/fMBT), GraphWalker (graphwalker.github.io), SpecExplorer [33], TCases (github.com/Cornutum/tcases), etc.
- Academic prototype tools: ModBat [34], MoMuT [35], VERA [36], JTorX [37], Torxakis [38], TESTAR [39], etc.

## 4.3 MBT literature in practice and industrial contexts

While it seems that most of MBT literature have been studies which conducted in academic and lab settings, a subset of the literature are studies conducted in practice and industrial contexts. We review a few selected studies below.

An author with affiliation in both industry and academia reported his view of the state of the art and challenges of "industrial-strength" MBT [40]. The reported experience and opinions are based on a MBT tool named RT-Tester, developed by the author's team. The paper highlights the importance of selecting the right modelling "formalism" for the testing problem at hand, and the fact that development of models, properly, can prove to be a major hurdle for the success of MBT in practice. As a related factor, the required skills for test engineers developing test models are significantly higher than for test engineers writing conventional test procedures. Other key factors for successful industrial-scale application of MBT as reported in the paper were: tracing requirements to the model, and automated compilation of traceability data.

An experience report of introducing MBT in the context of a system named European Train Control System (ETCS), developed by a large European company, named *Thales* was reported in [41]. The authors argued that MBT is not applicable "out-of-the-box", and application of MBT in a given environment (industrial context) requires specific adaptations. The selected test model formalism was UML/OCL. Certain toolchain-specific model revisions had to be made, e.g., timed triggers had to revised in the UML semantics (meta-model). The team used Borland Together for formalizing and concretizing system models. The last sentence of the paper was: "*it seems like the industry may already be aware of the possible benefits of MBT but fears the issues and costs of its integration*".

Microsoft has been one of the companies from which many MBT papers have been published, e.g., [42-44]. A 2003 paper [42] authored by a test architect at Microsoft reported the obstacles and opportunities for MBT in Microsoft. The author reported that: "*Model-based testing can provide a tremendous increase in testing capability, but modeling technology must be integrated into everyday software testing. Small-scale pilot projects, readily available tools and tester education have made the migration to test generation easier at Microsoft*". The author and his team used five characteristics of innovations that can accelerate or impede adoption, from a well cited book on the topic:

- Relative advantage: is your innovation better than the existing method?
- Compatibility: does your innovation integrate with the existing method?
- Complexity: is your innovation difficult to understand?
- Trialability: is it easy for people to experiment with your innovation?
- Observability: are the benefits of your innovation easily visible?

The author then reviewed how each of those characteristics affected the promotion of MBT at Microsoft. According to the paper [42], as of 2003, more than 600 of Microsoft 5000 testers were involved in some form of MBT.

Several papers from Microsoft have also presented their success story with MBT of documentation and quality assurance of client–server and server–server protocols of Microsoft Windows [43, 44]. A Microsoft MBT tool named SpecExplorer was used in those studies. The project was a large-scale undertaking in MBT: More than 25 000 pages of documentation for over 250 protocols had to be thoroughly verified to ensure that they are accurate, so that developers can implement protocols from the information they contain. Application of MBT reflected an investment of over 50 person-years. In addition, a substantial time investment was made in tool development, based on a continuous feedback loop from the test-suite development process into the SpecExplorer development team. According to statistical analysis, MBT resulted in a 42% productivity gain when compared with traditional test suites in a site where similar numbers of requirements were verified.

An interesting "voice of evidence" paper about MBT was published in IEEE Software in 2008 [8], which was based on systematic literature review (SLR). The authors argued that a rich body of experiences has not yet been published on all the SE techniques that researchers have proposed, including MBT. In fact, by some estimates, the techniques for which we do have substantial experience are few and far between. Thus, our current paper is a suitable evidence/experience paper aiming to address that gap. Based on their experience, the authors reported that: "*most developers [still] don't view MBT as a mainstream [testing] approach*" [8]. The study reported a "*serious lack of evidence*" in usefulness of different MBT approaches [25], and that many publications on MBT provide *only toy examples* without proper comparison with other approaches. The SLR divided the MBT studies into five categories: speculation, example, proof of concept, experience/industrial reports, and experimentation. UML-based MBT models were by far the most widely used formalisms. Furthermore, since applying MBT has non-trivial costs, the associated cost-benefits should be carefully analyzed when considering MBT, a topic referred to as "value-based" SE [24]. The study discussed this issue by stating: "*it's risky to choose an MBT approach without having a clear view about its complexity, cost, effort, and skill required to create [develop] the necessary models*" and that: "*Evidence on these topics could be a useful step in determining whether wider deployment of MBT approaches to different domains is worthwhile*".

### 4.4 MBT body of knowledge in the Formal Methods community

Researchers in the Formal Methods community have also done a large number of works on MBT since a few decades ago, e.g., see a short survey paper [45]. For example, a MBT approach using Labelled Transition Systems (LTS), which is a formal method notation, was presented in [46]. An approach for inferring finite-state machines (FSM's) was presented in [30], and those FSM's can later be used in MBT. Some fundamental work was done by Nicola and colleagues on testing "equivalences" [47] which have been highly cited in follow-up MBT studies. Various MBT tools have also been proposed by the Formal Methods community, e.g., [37, 38].

## 5 PHASES AND ACTIVITIES OF THE MBT TEST-AUTOMATION PROJECT

As the "core" of our work, we present the phases and activities of our MBT test-automation project, which include the followings. We first present our MBT test-automation strategy [15], which itself consists of: (1) how we selected the "right" test automation tool; (2) how the test models were designed; and (3) To enable full automated execution of MBT models, there is a need to development some type of "glue" code.

One of our goals in the project was to measure requirements coverage and ensure requirements traceability, which we will also present next. We will also report some results from execution of MBT test suites. Last but not the least, we will discuss briefly about development of an MBT coverage tool, that we saw the need for, during the project.

### 5.1 Test-automation strategy

For any test automation project, having a proper strategy is vital [15]. Such a strategy should include the following aspects: choosing the right test automation tool(s) [48], and how to develop the test scripts to ensure their quality (e.g., maintainability) [17]. We discuss next how we approached each of those issues in our MBT project.

#### 5.1.1 Choosing the right test automation approach and tool (ARQ1)

"*Selecting the right tool for the right purpose [in MBT] is a key to success*" [25]. A large number of MBT tools exist, either as commercial tools, open-source or academic prototype tools. As it has been reported in other areas of software testing, e.g., [49, 50], the choice of test tools often play an important role in success or failure of test automation endeavors.

A Google search for "model-based testing tool" would return the names and links to at least a few hundred such tools. For any test engineer, including us, choosing the "right" MBT tool is thus not trivial. For making such a choice, one would also experience the "paradox of choice", a phenomenon referred to as "the agony of choice", in an MBT book [26]. This phenomenon has also been reported in other areas of SE (bit.ly/SoftwareEngPOC). While there are comparative studies such as [7], we felt there was a lack of practical / pragmatic / "in-depth" studies comparing MBT tools, a need which we believe should be addressed by future studies.

To choose the right tool, we did not have the time resources to consider and exhaustively compare "all" the MBT tools out there, since there are simply too many tools. As discussed in Section 4.2, we relied on survey papers in this area [6, 7, 31] and also our exploratory Google search (relying on Google's PageRank) to hand-pick a manageable list of tools. The following tools were those that appeared in our candidate list:

- Commercial tools: TestModeller (testmodeller.io), TestOptimal (testoptimal.com), Tricentis Tosca (tricentis.com)
- Open-source/free tools, made in industry: SpecExplorer [33], GraphWalker (graphwalker.github.io), NModel [51], TCases (github.com/Cornutum/tcases)
- Academic prototype tools: ModBat [34], MoMuT [35], CrawlJax [52]

For choosing the right testing tools in general, many practitioners have offered experience-based heuristics. A Grey-Literature Review (GLR) done in 2017 [48] synthesized the heuristics reported in 53 blogs and white papers. The study presented 17 different criteria for choosing the right tool under three categories: (1) test-requirements and test-environment factors, (2) test-tool technical factors; and (3) test-tool non-technical factors. The five top criteria (of those 17) were: (1) the tool matching the test requirements, e.g., type of SUT (for us, this was web/mobile apps), (2) tool being fit to the operating environment, e.g., "right level" of model abstraction, test team's expertise; (3) tool's cost, (4) usability, and (5) availability of support for the tool.

We conducted a pilot phase in which we reviewed each tool's website to get familiar with its features and its modeling semantic. We also downloaded and tried the tool on one of the company's web applications (*Testinium*, testinium.com) to be able to assess it w.r.t. the above five criteria. To make the evaluation of the above criterion #2 (tool being fit to the operating environment) precise, we divided it into two parts: (2a) "right level" of model abstraction, and (2b) learnability of the tool, given our test team's expertise. Furthermore, in discussion with test engineers in the company (Testinium A.Ş.), we identified two of the criteria (1 and 2a) as "essential", i.e., if a tool fails any of them, it is out of the consideration. Results of our evaluation of the 10 above MBT tools w.r.t. our evaluation criteria are shown in Table 2.

**Table 2- Assessing a set of 10 MBT tools w.r.t. our evaluation criteria**

| Tools | Criteria | | | | | |
|---|---|---|---|---|---|---|
| | 1-Matching test requirements- Essential | 2a-Right level of model abstraction- Essential | 2b-Learnability | 3-Tool cost | 4-Usability-Essential | 5-Support |
| **TestModeller** | | Exhaustive activity diagram, resulting in repetition of nodes (youtu.be/nctAQHsmjpI) Failed | | | | |
| **TestOptimal** | Supports web/mobile apps, but not specific for them | Web page UI flow diagram | Reasonable | Has a free Community version. Paid professional version. ⇓ | Reasonable (our own usage, and youtu.be/cVlyV7eYbq8) | Has a Q/A page, with very few activities ⇓ |
| **Tricentis Tosca** | Seems like a test-data management tool (youtu.be/f6aBpa95kLc). While the introduction on its website mentions MBT, support for MBT is very limited. Not possible to design cycles and complex flow/edge structures Failed | | | | | |
| **SpecExplorer** | Support for MBT of web/mobile apps seems very limited. Most of focus is on API and unit testing. Failed | | | | | |
| **GraphWalker** | Specific for web/mobile apps | Web page UI flow diagram | Reasonable | Free open-source ⇑ | High ability to monitor the model during execution live (elements highlighted) ⇑ | Has a Wiki and Forum (active discussions) ⇑ |
| **NModel** | | Test model is in a programmatic format, instead of visual diagrams (doi.org/10.1007/978-3-642-05031-2_14) Failed | | | | |
| **TCases** | Support for MBT of web/mobile apps seems very limited. Most of focus is on test-case design for input space exploration. Failed | | | | | |
| **ModBat** | Focus is on API testing. No support for MBT of web/mobile apps. (fmv.jku.at/modbat) Failed | | | | | |
| **MoMuT** | Focus is on embedded system testing. No support for MBT of web/mobile apps. (momut.org) Failed | | | | | |
| **CrawlJax** | It produces as output a state-flow graph of the dynamic DOM states and the event-based transitions between them. Focus is not on GUI testing of web apps Failed | | | | | |

As we can see in Table 2, only two tools (TestOptimal and GraphWalker) have pasted the filtering. After a careful investigation, and as the assessments of these two tools in in Table 2 show, we selected GraphWalker, due to the following rationale: (1) it fit our needs, and was open-source, thus we could also modify it to meet our purpose, if we wanted to; (2) its modeling semantic was simple, light-weight and pragmatic; and (3) since it is open-source, we did not have to worry about availability of support for the tool. Furthermore, many of the academic tools were mostly prototypes, thus were not *production-ready* for our purpose, and most were based on heavyweight modeling formalisms. Furthermore, we found that several case studies using *GraphWalker* have been shared by other test engineers, e.g., testing an information kiosk (panel) software in New York's subway (bit.ly/MBTGuidingTestingDecisions) and also for testing games (bit.ly/MBTofAGameEngine), thus showing its applicability and usefulness in practice.

**Lesson learned**: We empirically observed that choosing the "right" MBT tool from amongst the very large pool of available MBT tools, for a given industrial testing context and project, is challenging and not trivial. This validates the empirical evidence reported in many academic and grey literature sources, e.g., [25, 48]. We found that, as also reported in many other resources, selecting the "right" tool for the "right" purpose in MBT is a key to success. Even if a team has the expertise and knows which MBT technique to use, but if the tool is not "right", succeeding in MBT will be less likely. We found the guidelines of a Grey-Literature Review (GLR) [48] in this topic useful as they helped us choose the right tool.

Based on how *GraphWalker* works, we designed our MBT approach as shown in Figure 2. Test engineers uses the system requirements to design the test models, a form of activity diagrams showing the UI flow across different pages of a web application under test. Test engineers should also develop the Selenium Java code to "implement" the action of each node/edge in the MBT test models. MBT test models are then executed using the chosen test tool (*GraphWalker*), which uses the developed Selenium Java code to exercise (call) the front-end of the web application under test, and that communicates with the back-end. Test outputs are recorded, logged and returned to test engineers by the chosen test tool (*GraphWalker*). We discuss each of the steps of Figure 2 in more detail in the next sections.

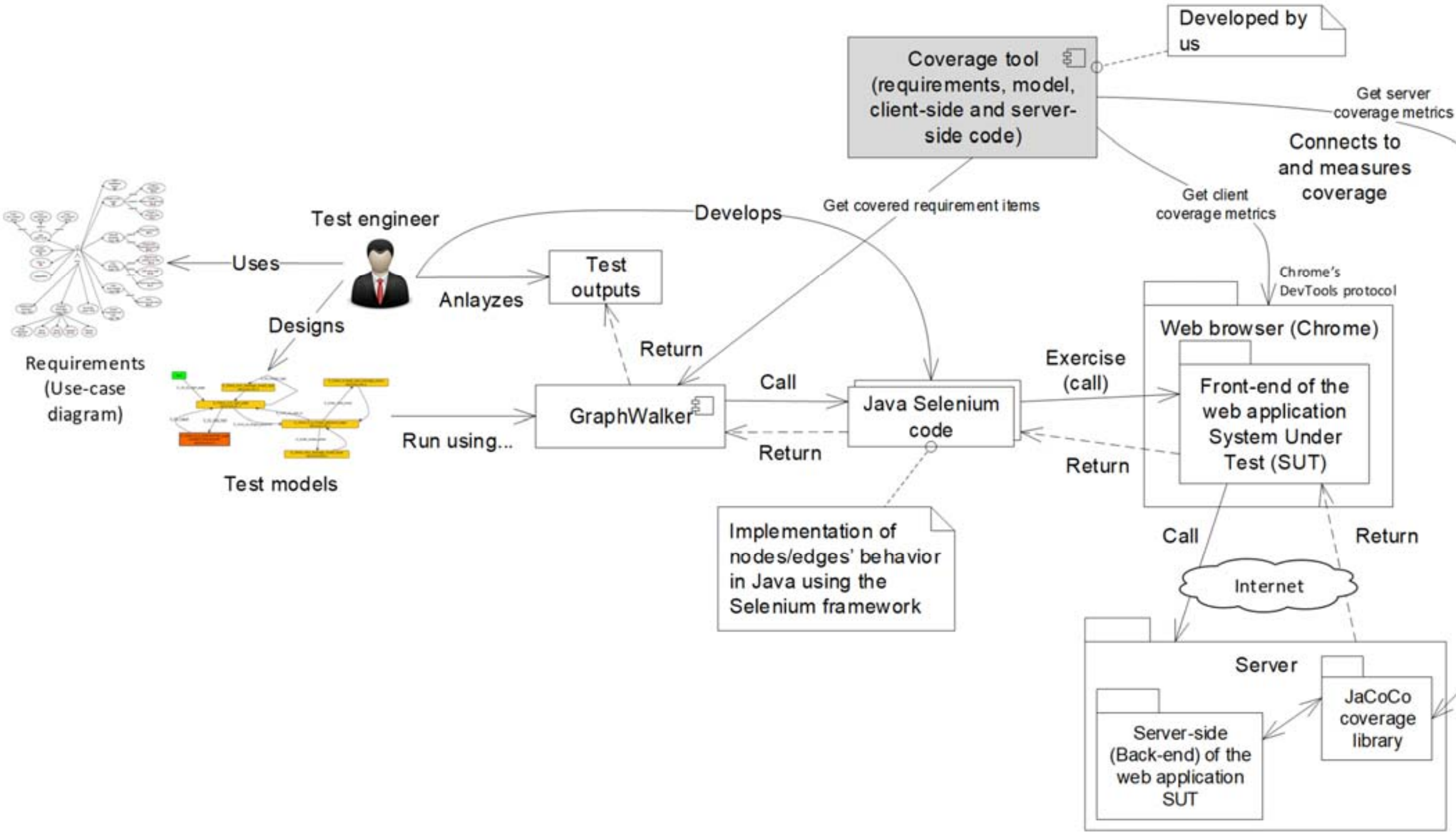

**Figure 2- An overview of our MBT approach**

**Lesson learned**: When introducing MBT to a company for the first time, a lightweight MBT tool/approach is advisable, especially when there exist success stories from other practitioners that have successfully used a given MBT tool in other industrial contexts (companies).

### 5.1.2 How the test models were designed

Among important issues in conducting MBT are levels of abstractions and granularity in test models [26]. They directly impact how engineers should design the test models. Generally, one has to choose the "right" level of abstraction and this impacts the choice of MBT tool and approach. The modeling formalism, abstraction level and granularity, followed by our chosen MBT tool, showed to be practical and appropriate, for the context and domain at hand (web applications).

Let us continue with concrete examples from one of our actual SUTs: *Testinium* (testinium.com), the flagship test tool of the company, which is a web-application gateway (wrapper) on the *Selenium* test framework and provides test-management features and testing on the cloud. Figure 3 shows two screenshots from the SUT: the login screen and the "dashboard" (main page) shown just after login. Essentially, we used the MBT approach to *test this test tool*. Our goal was to deploy MBT extensively for this large SUT and use the knowledge and expertise that we and our test engineers would learn in the process to increase the capacities of the test team in testing of the many SUTs provided by the company's clients.

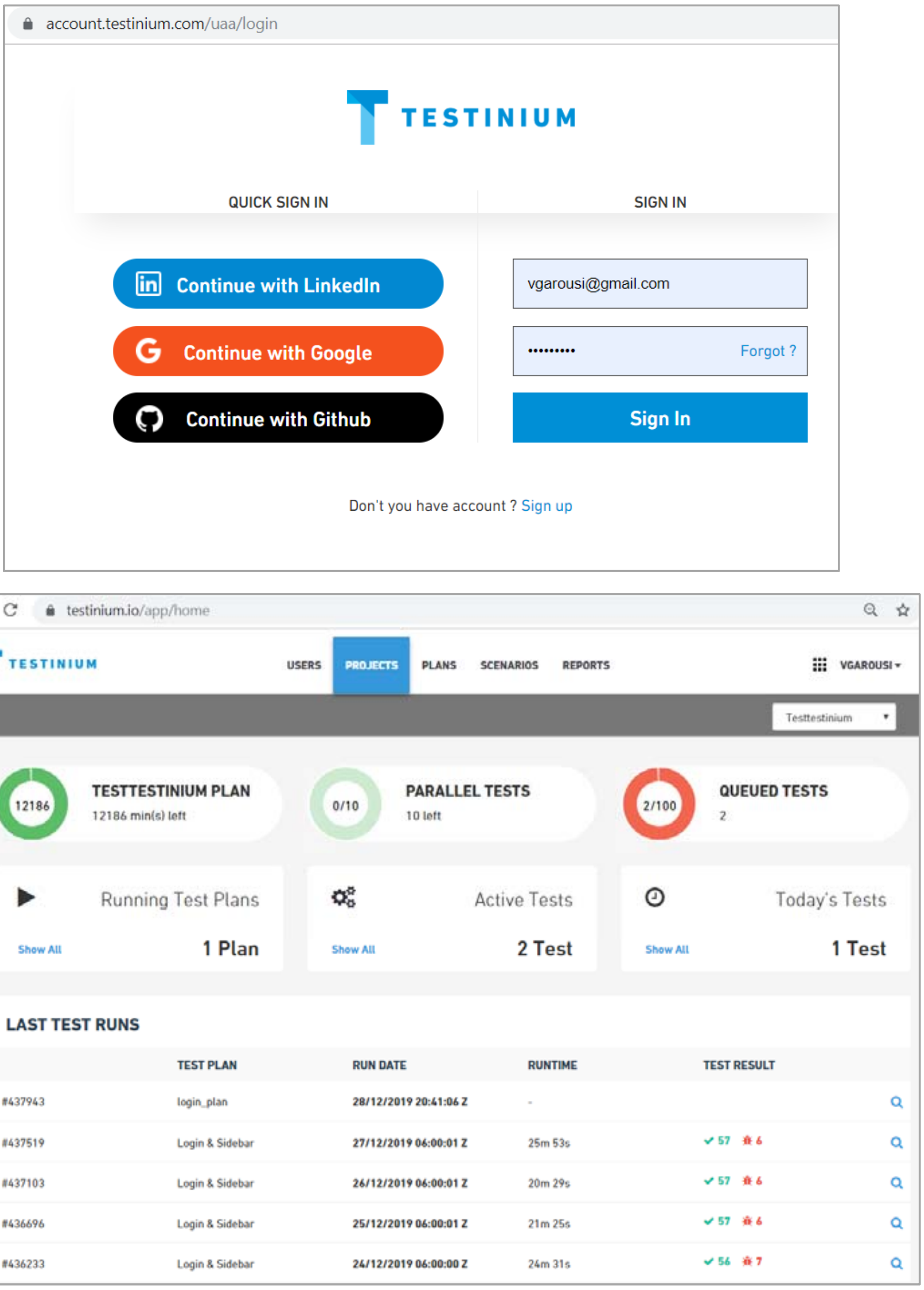

**Figure 3- Screenshots from the SUT: Testinium**

We show in Figure 4 two test models designed for testing the above two pages. In the modeling semantic of the tool, each edge corresponds to an action (stimulus), e.g., `e_click_signin`, and each node corresponds to one or more verifications (to be developed using "assert" functions in Selenium Java code), e.g., `n_verify_in_forgot_password_page` in Figure 4. In this MBT approach, test models are lightweight UML activity diagrams, and are essentially the webpage flow-graphs of the web application under test. The formalism supports definition of certain nodes as "shared" nodes (shown with orange color in Figure 4), which allow breaking down the entire system to several models. When visiting a shared node, the tool jumps to any node that has the same tag (they are like function calls). For the web applications domain, this lightweight notation can be considered a domain-specific test modeling language.

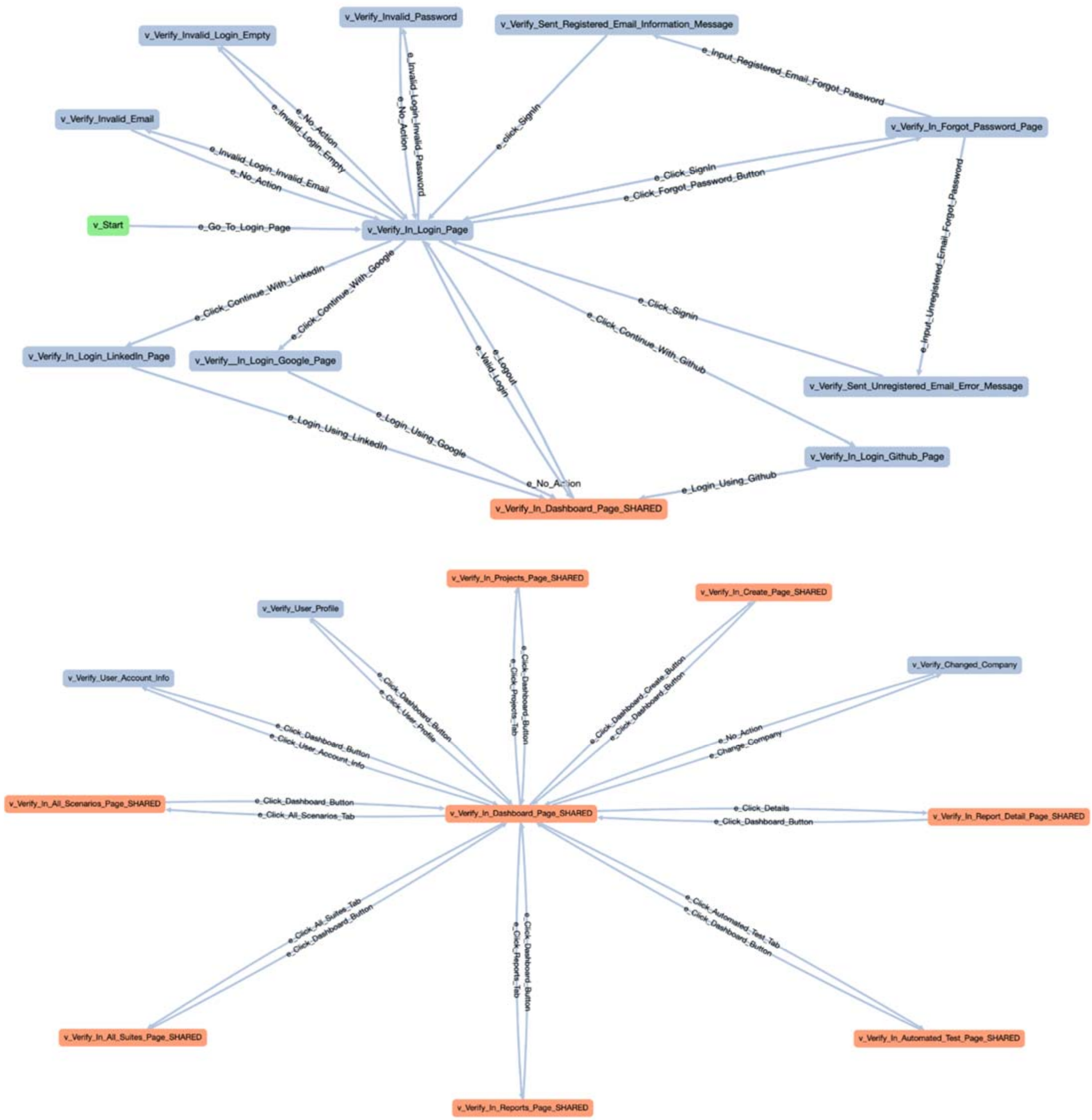

**Figure 4- Two MBT test models for the SUT: the MBT models of the *login* and *dashboard* pages (shown in Figure 3)**

The MBT models could be, in principle, developed either manually or automatically, i.e., reverse-engineering of the web GUI, also called GUI "ripping" [53]. While we have had some prior experience using some of GUI ripping tools [54], and we actually tried the possibility of using that approach, we soon noticed that one disadvantage is getting very large models with many details (clicking on every possible link in web pages), that later would require test engineers to spend a lot of effort to prune ("clean") them to make them executable in MBT tools.

After some evaluations with our team-members, and since we found that developing test models manually did not take too much effort and, in fact, did provide various “side” benefits (discussed next), we decided to develop the test models manually. Test engineers actually benefitted from and liked the effort put into developing test models, since it was quite a valuable learning experience for them to better understand the SUT and their test approach, an observation also reported elsewhere [26]. Also like other studies [25], we observed that "*testers working with MBT have increased motivations and are eager to learn*".

**Lesson learned**: Even if the MBT models may be developed semi-automatically by reverse-engineering them from the web SUT, we however found that manual development of MBT models by test engineers provided various "side" benefits, e.g., valuable learning experience, increasing motivations and interest of test engineers in test automation. Also note that, if we use tools to reverse-engineer the MBT models, the huge effort to prune (clean) them to make them executable in MBT tools often overweighs the cost of developing them from scratch manually.

One important point is about design best-practices for models. When developing the test models, we used *GraphWalker*'s online guidelines (graphwalker.github.io) and the chapter "*Good MBT modeling practices*" in an MBT book [26] to ensure high-quality design for test models, which could be called "model design patterns", similar to object-oriented (OO) design patterns. For example, test models should be designed in a way to be understandable and maintainable. Aside from the above sources, we found only a few sources in peer-reviewed and grey literature on this topic, and thus we think there is a need for more research on this topic in future.

**Challenge**: We observed a general shortage of knowledge and resources on best practice and "design patterns" for designing MBT models. We thus recommend more research and investigations on this very important topic by researchers and practitioners in future.

Last but not the least in this section, we discuss the size metrics of the MBT test suite. Since the SUT (Testinium) had 18 distinct UI pages, our MBT test suites for the SUT resulted in 18 test models (two of them are shown in Figure 4). Altogether, those 18 test models had 177 nodes and 260 edges. We have made all the MBT models and artifacts of this SUT available as open-source in: github.com/vgarousi/MBTofTestinium.

### 5.1.3 Development of nodes/edges' behavior in Java using the Selenium framework

As shown in Figure 2 (our MBT approach), testers need to provide the behavior of nodes/edges in Java using the Selenium framework. Once we ensured that our test models are properly designed (we did a few round of peer reviews), we developed the Java test-code. For example, for the edge `e_valid_login` in Figure 4, we developed the Java test code shown in Table 3. In this example Selenium Java code, to conduct a valid login, the username and password fields are first located. Then, a correct combination of username and password values are entered in those fields. To find the HTML button for the "Sign in", a CSS selector path is given. The sign-in button is finally clicked programmatically.

As per our observations, the relatively-short Java methods implementing nodes/edges' behavior were quite trivial to develop and we did not notice any noticeable challenges.

**Table 3- Java Selenium code implementing the behavior for edge `e_valid_login` in Figure 4**

```
public void e_valid_login() {
   WebElement userNameElement =
       methodsPage.findElement(By.id("username"));
   userNameElement.clear();
   userNameElement.sendKeys(email);
   WebElement passwordElement =
       methodsPage.findElement(By.id("password"));
   passwordElement.clear();
   passwordElement.sendKeys(password);
   methodsPage.findEleminrent(By.cssSelector(
     "input[class$=\"login-page__submit-btn\"][value=\"Sign In\"]")).click();
}
```

We observed that, the chosen modeling semantic provided (in a sense, "enforced") a suitable "separation of concerns" (SoC) (design pattern) [55] in a way to make the test code modular and helped test engineers clearly know what to develop for each Java method (for example the above method). Also, each method was only a few lines of code, which we think is a best-practice on its own, conceptually similar to the following OO recommendation: "*Small methods are a hallmark of OO thinking*" (bit.ly/OOPrinciples).

Test-code development was incremental and test engineers would run the model after developing several methods to test the test suites, and make corrections if necessary. We used other test patterns when developing test code, e.g., "Page Object" pattern as seen as `methodsPage` in the code listing above. To ensure quality of test code, we also conducted peer reviewing.

Thus, chances of having defects in the test suites were slim, and in case of observing issues, the team was able to quickly find and resolve them.

**Lesson learned**: The modeling semantic of the chosen MBT tool provided (in a sense, "enforced") a suitable "separation of concerns" (SoC) (design pattern) in a way to make the test code modular and helped test engineers clearly know what to develop for each Java method. Also, each method was only a few lines of code, which we think is a best-practice on its own, conceptually similar to the following OO recommendation: "*Small methods are a hallmark of OO thinking*".

**Advice**: Even when using a lightweight MBT tool/approach, there is work that needs to be done manually by test engineers. However, such work was not more difficult or time consuming than writing test cases by hand, nor it required any special in-depth training to learn to use a tool like GraphWalker.

### 5.2 Requirements coverage and requirements traceability

One of the work packages in our original project plan (testomatproject.eu) necessitated measuring requirement coverage in design and also execution of test suites and also incorporating test-requirement traceability. Since our context is an agile context, there were no formal pre-written requirements documents for any of the SUTs, including the SUT discussed in this paper (Testinium). We did lightweight reverse engineering of use cases for the SUT (Testinium) based on the actual implemented system, as shown in Figure 5.

The MBT tool has a simple but effective feature for requirements coverage and traceability, as shown in Figure 6. Using the step labels in the description of each use-case (such as R1.1), we labeled each node of the test model accordingly, and in this way, at end of each test execution, the MBT tool provides the ratio of requirements coverage, as a percentage value.

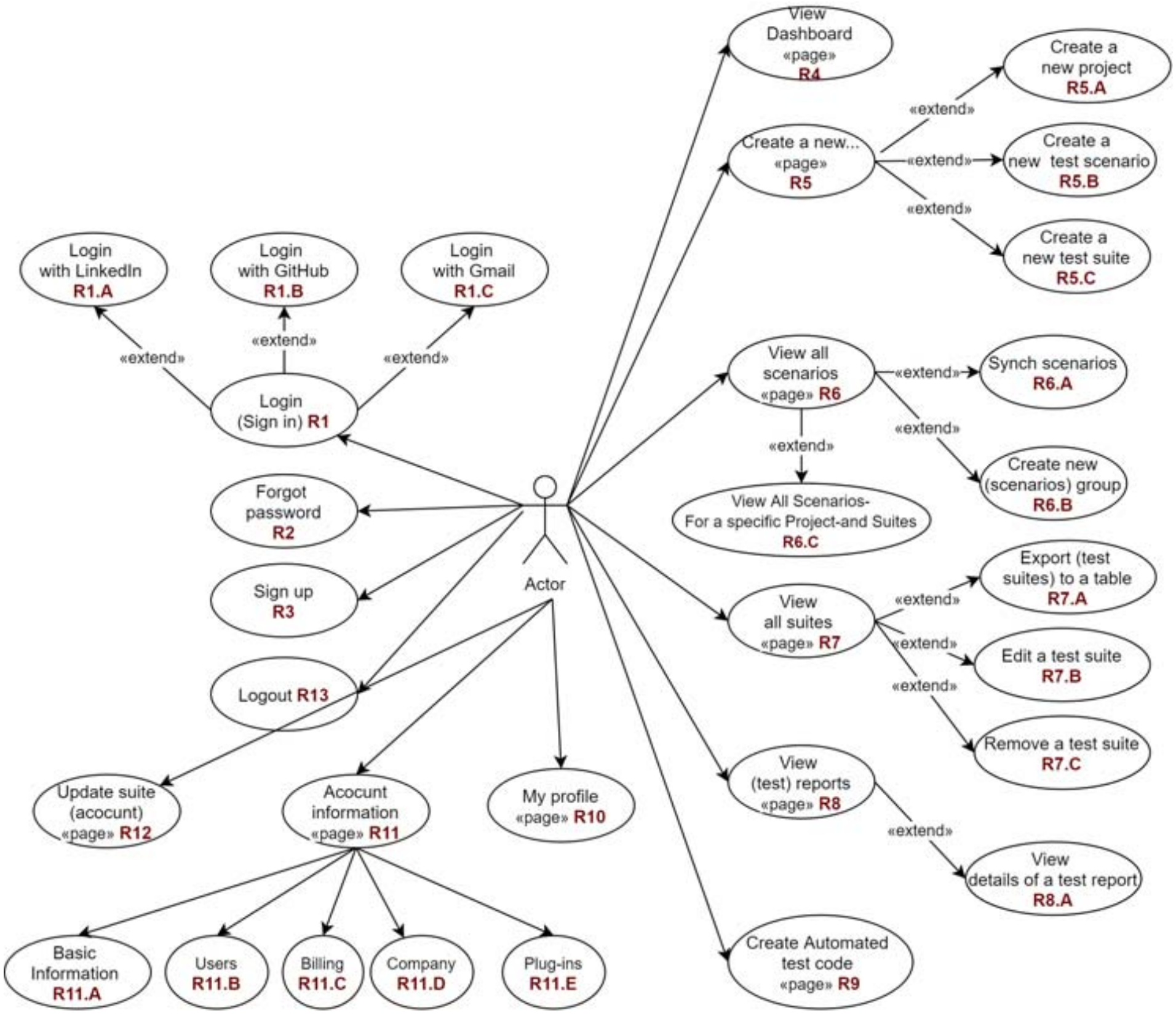

Use-case ID: R1
Use-case Name: Login (Sign in)

| Steps | |
|---|---|
| **User does...** | **System should do...** |
| User accesses the "Login" page | R1.1-System shows the "Login" page |
| User enters VALID login info and clicks "Login" | R1.2-System shows the "Dashboard" page |
| **Alternative paths:** | |
| User enters INVALID login info and clicks "Login" | R1.3-System shows an error message and stays in the "Login" page |

**Figure 5- The SUT use-case diagram, and an example use-case description**

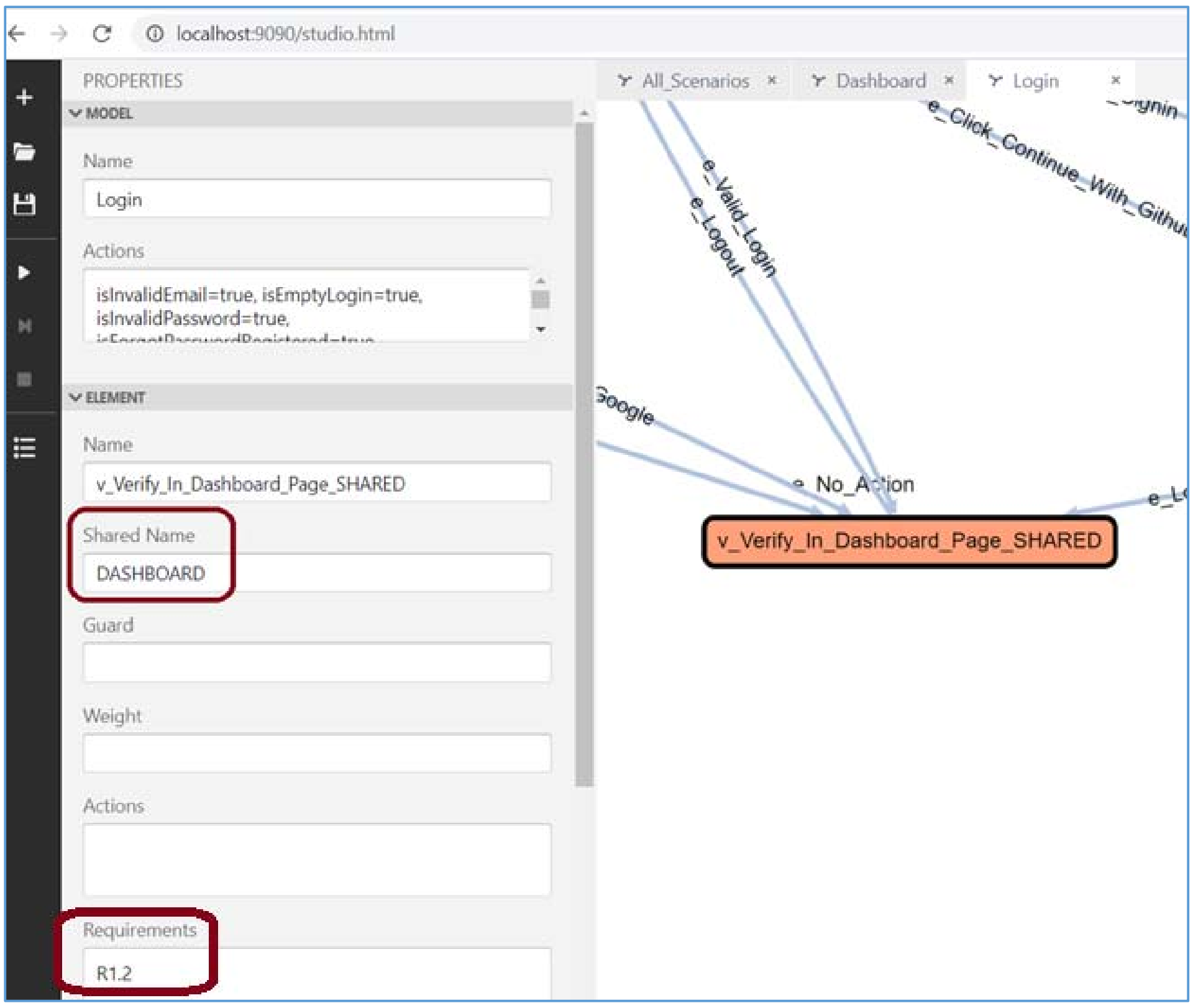

**Figure 6- Assigning a given node to a requirements item in the chosen MBT tool**

### 5.3 Video demos and the project artifacts

For interested readers, we have recorded several video screencasts of the MBT test executions and have posted them on YouTube (bit.ly/VideosMBTTestinium). Also, to help other practitioners review and learn from our MBT project, we provide the entire test artifacts (test models and Java codes) of the Testinium SUT, open source, in a GitHub repository (github.com/vgarousi/MBTofTestinium). We have also posted the archived version of the MBT test-suite code and one of the videos in a permanent location with a Digital Object Identifier (DOI) [56].

### 5.4 Execution of MBT test suites

Once we utilized best practices to iteratively design and develop the MBT test models and the required test artifacts (test code to implement nodes/edges' behavior in Java using Selenium), we could then start running the full MBT test suite on the production SUT (Testinium). We soon decided to embed the MBT execution in the company's Continuous Integration (CI) pipeline, which would run at least once every night and report the results. We show in Figure 7 an email screenshot from the nightly auto-scheduled MBT executions in the CI pipeline. The two attached TXT and Excel files are detailed logs of test executions, i.e., paths, nodes and edges covered in the test run. Later in the paper, in Figure 9, we show a partial snapshot from the output test log, an Excel file automatically generated by the extended reporting engine that we have added to the MBT tool (GraphWalker), and emailed automatically, as shown in Figure 7.

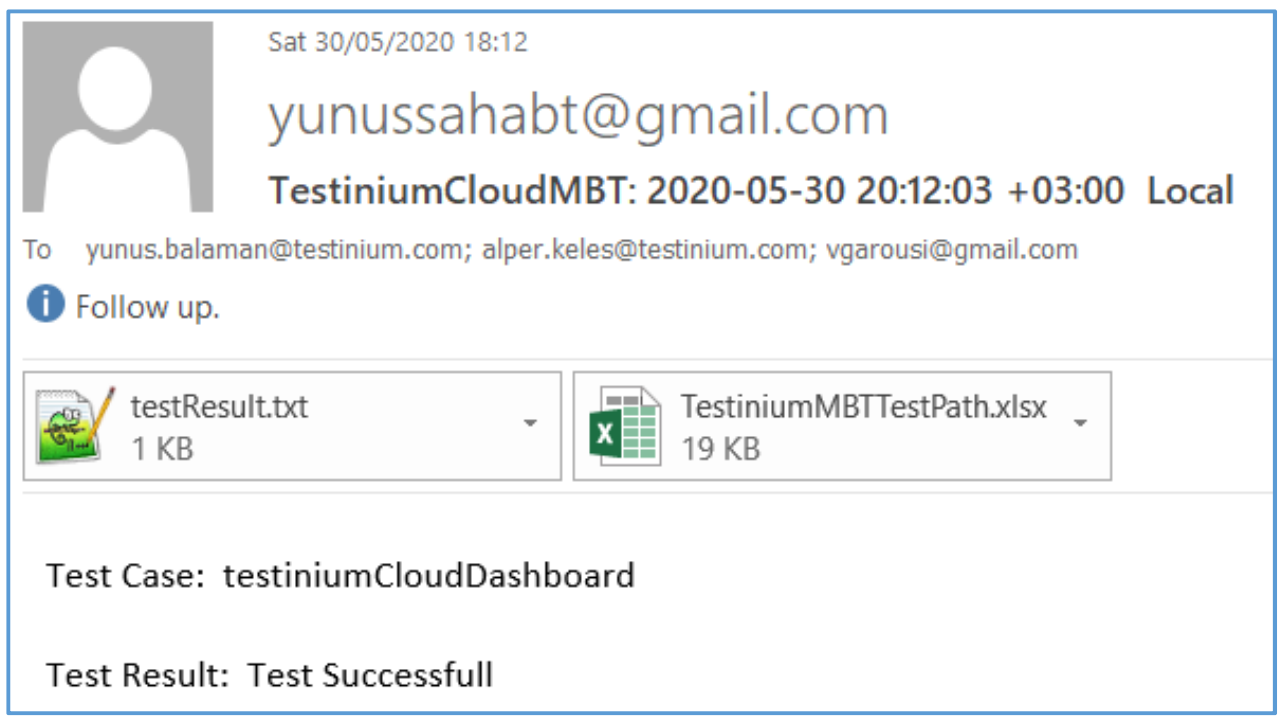

**Figure 7- An email screenshot from the nightly auto-scheduled MBT executions in the Continuous Integration (CI) pipeline**

**Lesson learned**: We found the practice of including the MBT execution in the Continuous Integration (CI) pipeline to be a good approach, as it would execute automatically every night and report the results.

For execution of MBT test suites, the other important aspect was setting the MBT tool parameters and configurations. The chosen test tool (GraphWalker) provides a large number of parameters for designing and running a given MBT Test suite (all details can be found in the tool's website and online documentation). Two main parameters, worth mentioning, are the level of model coverage that the test engineers want to cover the models for test-case generation and execution and the type of graph traversal strategy. GraphWalker documentation phrases this as follows: '*Path [test-case] generation consists of two parts: "how to cover?" (generators) and "what to cover?" (stop conditions)*'. A generator is an algorithm that decides how to traverse a model. Four graph traversal algorithms ("generators") are supported by GraphWalker, as of this writing[1]:

- Random: Navigate through the model in a completely random manner, also called "Drunkard's walk", or "Random walk". This algorithm selects an out-edge from a vertex by random, and repeats the process in the next vertex.
- Weighted random: Same as the random path generator, but will use the weight keyword when generating a path. The weight is assigned to edges only, and it represents the probability of an edge getting chosen.
- Quick random: Tries to run the shortest path through a model, but in a fast way. This is how the algorithm works:
  - Choose an edge not yet visited by random.
  - Select the shortest path to that edge using Dijkstra's algorithm
  - Walk that path, and mark all the executed edges as visited.
  - When reaching the selected edge in step 1, start all over, repeating the above four steps
  - The algorithm works well for very large models, and generates reasonably short sequences. The downside is when used in conjunction with extended finite-state machine, the algorithm can choose a path which is blocked by a guard.
- A* (A-star): Will generate the shortest path to a specific vertex or edge.

We should remind that the above four algorithms are among the classical graph traversal algorithms and further details about them can be found in a typical graph theory textbook [57]. We have done some initial experimentation with some of the above, but to keep the complexity of our work in a manageable level, we have configured the daily MBT run to use the "Random" option. We plan to conduct in-depth studies by varying the choice of generators.

**Open question**: Any given MBT tool and approach (including the one that we selected), has various parameters and configurations to be set, like the ones above. An important open question is which sets of parameters are the best, or would provide better test outcomes, coverage, execution time, fault detection effectiveness, etc.? This raises the need for empirical studies on the choice of those parameters and configurations and also possibly some new Search-based Software Engineering (SBSE) [58] approaches to select the best options.

Another important aspect for MBT test execution is stopping conditions (criteria), a condition that decides when MBT test execution stops. The generator will generate a new step in the path until the stop condition is fulfilled. Nine different stopping conditions are supported by GraphWalker:

[1] https://github.com/GraphWalker/graphwalker-project/wiki/Generators-and-stop-conditions

1. Edge coverage: When, during execution, the percentage of traversed edges is reached, the test is stopped. If an edge is traversed more than once, it still counts as 1 when calculating the percentage coverage.
2. Vertex (node) coverage: When, during execution, the percentage of traversed states is reached, the test is stopped. If a vertex is traversed more than once, it still counts as 1 when calculating the percentage coverage.
3. Requirement coverage: When, during execution, the percentage of traversed requirements is reached, the test is stopped. If a requirement is traversed more than once, it still counts as 1 when calculating the percentage coverage.
4. Dependency edge coverage: When, during execution, all of the traversed edges with dependency higher or equal to the dependency threshold are reached, the test is stopped. If an edge is traversed more than once, it still counts as 1, when calculating the percentage coverage. The concept of "dependency edge" is actually more like operational profiles [59], i.e., putting weight values on edges.
5. Reached vertex: The stop condition is a named vertex. When, during execution, the vertex is reached, the test is stopped.
6. Reached edge: The stop condition is a named edge. When, during execution, the edge is reached, the test is stopped.
7. Time duration: The stop condition is a time, representing the number of seconds that the test generator is allowed to execute.
8. Length: The stop condition is a number, representing the total numbers of edge-vertex pairs generated by a generator.
9. Never: This special stop condition will never halt the generator.

Again, we would have liked to put more resources and experiment with various parameters, but for the time being, we decided used the Edge coverage=100% as the stopping condition, which we believe is a reasonable (and acceptable) stopping condition, at least for all the test engineers that were involved in this project.

With the above parameters, each full execution of the MBT test suite would take about 6 hours. As discussed in Section 5.6, we provide a glimpse of test execution in several YouTube videos (bit.ly/VideosMBTTestinium).

### 5.5 Development of an MBT coverage tool

Two classical approaches for assessing effectiveness and efficiency of any testing technique are: (1) its ability to detect defects (real or artificially injected defects, via mutation testing), (2) how much coverage is achieved during test execution; and test coverage can have many different forms, e.g., requirements coverage, code coverage and MBT model coverage.

As we were preparing and planning to evaluate the benefits and effectiveness of the MBT approach w.r.t. the second important aspect above (coverage), we searched for available coverage tools to apply in our context.

For measuring code coverage for web applications, one needs to measure both front-end (client-side) JavaScript (JS) and also back-end (server-side) coverage values. In our search for JS coverage tools, we came across many tools, e.g., the Istanbul tool (istanbul.js.org), the "Developer tools" (DevTools) protocol of Google Chrome (developers.google.com/web/tools/chrome-devtools). For assessing back-end (server-side) coverage, there are also various tools, depending on the server-side technology, e.g., the JaCoCo code coverage library (jacoco.org) for server applications developed in Java, xDebug (www.xdebug.org) and PVOC (github.com/krakjoe/pcov) for server applications developed in PHP. While all these tools are quite stable and popular for their purposes, our code-coverage need in our context was to gather and present both client-side and server-side coverage values in one user- (tester-) friendly output (e.g., in line charts), in a "live" manner (as a given MBT test suite was running), and would "connect" to our selected MBT tool (GraphWalker) seamlessly (without hassle). For such a requirement, we did not find any readily-applicable tool to work in conjunction with MBT for web applications.

On the other hand, in terms of showing the MBT "model" coverage, our selected MBT tool (GraphWalker) would only show the coverage values (how many edges and nodes have been covered) at the "end" of MBT test execution and not "during" test execution. In discussions with the test engineers in the company (Testinium A.Ş.), they mentioned to us that, for a test engineer, it is much useful to observe code and model coverage during MBT test execution, especially since such a test execution for a medium size SUT (e.g., Testinium itself) would take about 6 hours (as discussed in Section 5.4), and it is important get continuous regular feedback about test coverage which a test suite is running, not just at the end.

To meet all the above requirements, we decided to develop an MBT coverage tool to measure both model coverage as well as code coverage at front-end (client-side) JavaScript (JS) and also back-end (server-side) of the web application under test.

To develop such a tool, we had to choose a client-side and server-side coverage tools and "integrate" their outputs and show the results live visually. For model coverage, we used the API of our selected MBT tool (GraphWalker) to query the model coverage in regular intervals (e.g., every 5 seconds).

We named our new developed MBT coverage tool *MBTCover*. We have already made MBTCover open-source at: github.com/vgarousi/ MBTCover. Already, the tool has started to be downloaded by developers in the community.

We explain next some technical details about how we developed MBTCover. To get front-end (client-side) JS coverage values at runtime, we used the Chrome "Developer tools" (DevTools) protocol. To programmatically extract coverage live from DevTools at runtime, we use a library called Puppeteer (www.pptr.dev) which provides an Application Programming Interface (API) to the DevTools protocol.

To get back-end (server-side) coverage live at runtime, we used the JaCoCo code coverage library (jacoco.org). This was a suitable choice since the implementation language of the SUT in our running case (Testinium) was Java. Of course, for other SUTs which have been developed in other programming languages (such as .Net), other server-side code coverage technologies should be used.

Further details about our implementation of MBTCover can be found directly in its open-source code-base at: github.com/ vgarousi/MBTCover.

Two screenshots from the MBTCover tool are shown in Figure 8, in which the SUT is Testinium and the MBT suite is running. Two charts, developed in JavaScript (JS), are updating live every few seconds, which is an option chosen by the user, showing the front-end (client-side JS) coverage: (1) One chart shows the *cumulative* front-end (JS) coverage, meaning that the coverage calculation has been done based on the combined lines of JS covered in all the web pages of the SUT, reached so far, divided by the sum of all JS code lines; (2) The other front-end coverage shows the JS coverage % of the *current* web page, being tested by the MBT suite.

In the current implementation, MBTCover measures the coverage of all the JS files: all third-party JS libraries imported in a web page and also the customized JS files developed for the SUT. We plan to develop in near future a feature to select which JS files to instrument and measure the coverage for. In the screenshots, we can see that the cumulative front-end (JS) coverage has increased from mid-0% to above 50% and then back to mid-20% as the MBT suite continues execution and visits different pages of the web app SUT. The reason for the fluctuation (up and downs) is that different web pages of the SUT use (reference) different JS files with different Line-of-Code (LOC) sizes and also those different web pages use (call) different amounts of JS LOC. Thus, the cumulative JS coverage would fluctuate as we can see in the screenshot.

In Figure 8, the other front-end coverage chart showing the JS coverage % of the current web page also provides valuable information, as we can see the extent of JS code coverage in the current page, as being tested by the MBT suite. For example, the MBT execution of Testinium starts with the Login page (Figure 3) and then moves to the Dashboard page (the first orange and the second yellow chart lines correspond to hose two pages). As expected, the current web-page coverage chart resets to the value of 0% in each page and then *grows* up to a certain level, until the web page changes as the MBT suite is commanding the SUT. The test engineer can see live the extent of coverage in the current page and take actions if they decides to, e.g., if the coverage is low, they can investigate why most parts of the included JS files have not been covered, and they may decides to add more test paths to the MBT models, etc.

As shown in Figure 8, another chart shows the back-end (server-side) coverage of the web application under test. In the current version of MBTCover, we are only showing one chart for server-side coverage, i.e., the cumulative code coverage. Of course, server-side coverage would depend on the technology and programming language of the SUT files developed to run in the server. For Testinium, this was the Java language. As we see in Figure 8, the server-side coverage value starts from about 10% as the MBT suite starts execution and of course, since it is a cumulative percentage value, it will either stay constant or go up. The server-side coverage value then slowly goes higher up to about 12% and then there is a quick jump to about 27% and then higher. That jump was due to the SUT going to a certain page ("Create Page" feature of Testinium) which exercises a large amount of the Java code-base on the server, thus leading to sudden increase in server-side coverage.

Furthermore, in discussions with the test engineers in the company (Testinium A.Ş.), they have mentioned to us that it would be useful/interesting to observe "cumulative" coverage values while testing. We have developed the MBTCover tool to work in this way, e.g., part (1) of Figure 8 shows the "cumulative" server-side coverage. Also, part (4) of Figure 8 also shows the "cumulative" client-side coverage. The reason why the coverage values in part (4) drop down almost mid-way in the shown chart is that new JS files (libraries, to be precise) have been imported in the HTML / JS files of client-side code of Testinium in the new pages of the SUT that is being tested by the MBT suite. Therefore, the JS coverage calculation (which is based on the formula: num_of_covered_JS_lines / total_JS_lines) would give a lower value since the "divisor" value in the formula increases suddenly.

Just like other aspects of our work in this project (designing the MBT models and the Selenium code), our software development process has been iterative and Agile. In iterative development of MBTCover, we saw the need to also show, in the GUI of the tool, several important informative (useful) statistics, as shown in Figure 8, which include: (1) number of test models reached so far, (2) number of nodes covered so far, and (3) number of nodes executed so far. Note that there is a difference between the last two mentioned items since the form is the node coverage of MBT models, while the latter is the number of nodes which have been executed and a given node could be counted more than once. Without these live metrics, the tester has to wait until the end of MBT execution (and that could take up to 6 hours for our test suite), to see the outcomes.

**Lesson learned**: While there are numerous coverage tools for conventional test automation, e.g., for xUnit frameworks, to our surprise, there were no off-the-shelf readily-applicable coverage tools to work in conjunction with MBT for web applications. Thus, practitioners should be aware of this, if they plan to measure coverage in conjunction with MBT. The MBTCover tool, that we have developed, works with the chosen MBT tool (GraphWalker), thus it can be helpful.

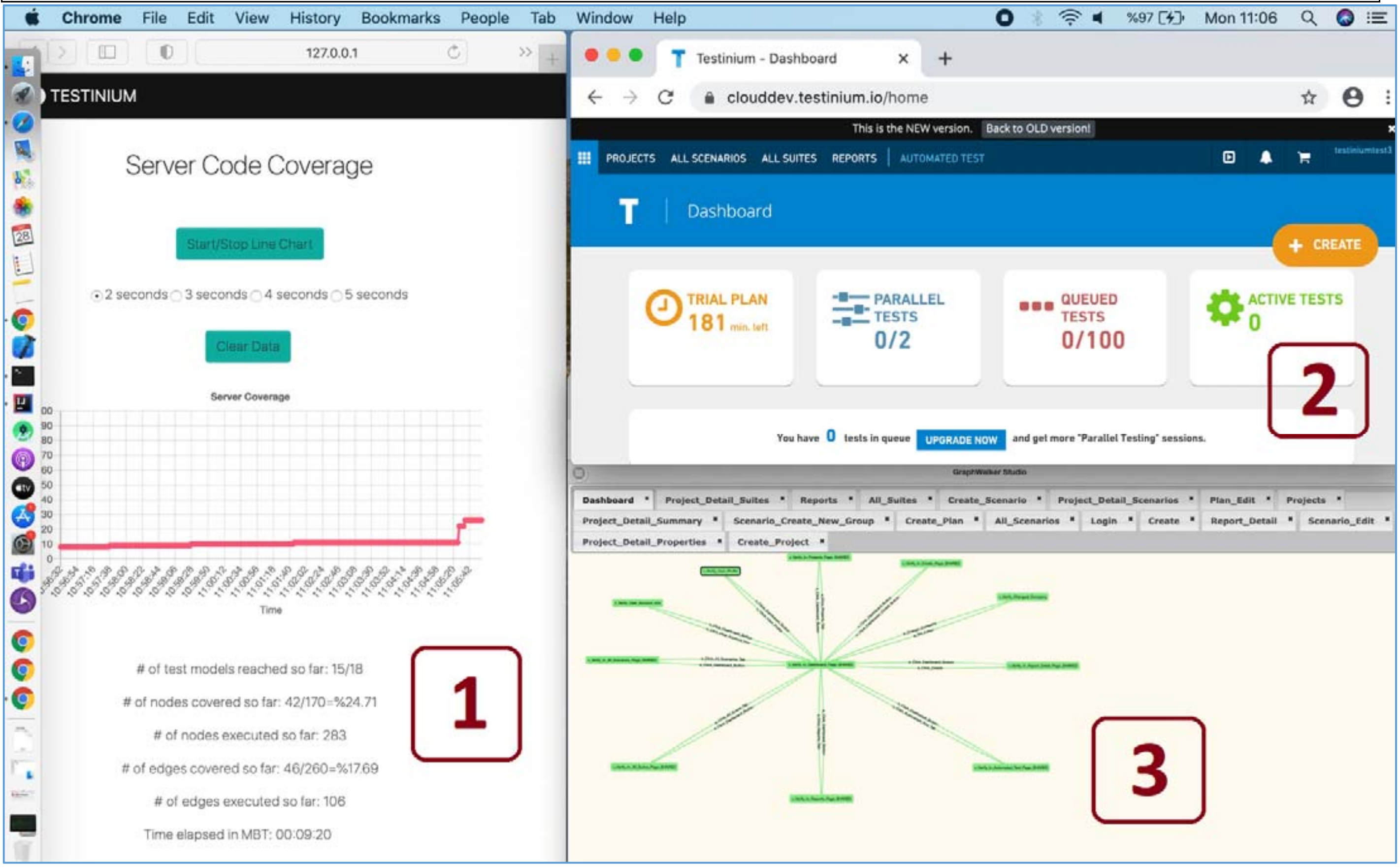

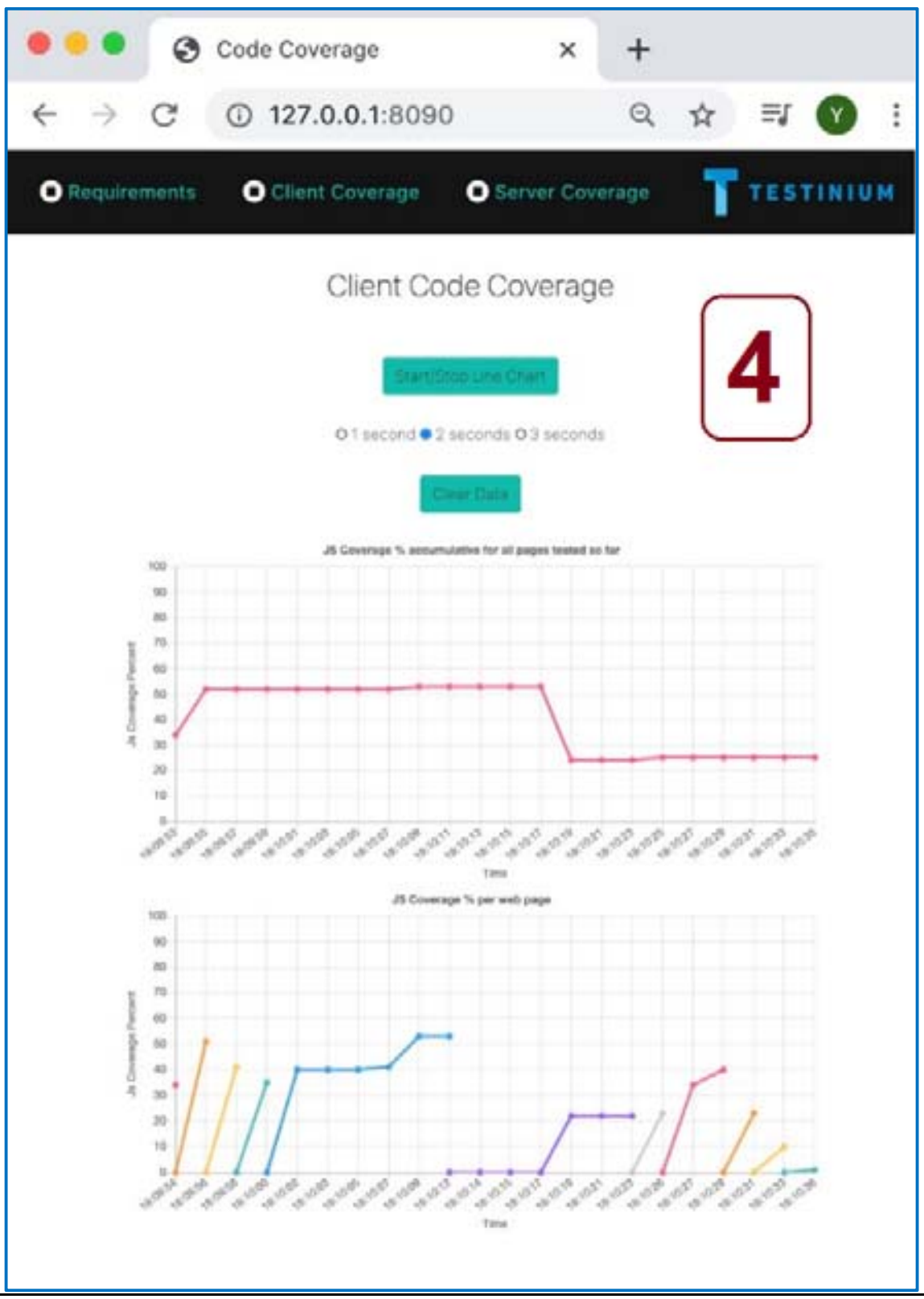

**Figure 8- Screenshots from the MBT coverage tool (MBTCover) that we have developed, as part of the project. (1) MBTCover: Server-side coverage; (2) SUT: Testinium; (3) MBT execution tool: GraphWalker; (4): MBTCover: Client-side coverage**

## 6 EMPIRICAL FINDINGS GATHERED SO FAR IN THE PROJECT

As discussed in Section 3, we derived in the planning phase of our project three Action-Research Questions (ARQ), two of which were:

- ARQ2: *What benefits does the MBT approach provide in the industrial context?*
- ARQ3: *Which challenges and questions did we face in the MBT project (so far) and how can they be addressed?*

During the phases and activities of the MBT test-automation project (Section 5), we conducted the measurements to be able to assess these two ARQs, and we report the empirical findings next.

### 6.1 Evaluating the benefits of the MBT approach (ARQ2)

To assess the benefits of MBT in our project, we identified its benefits and (positive) impacts compared to the previous test approach used in the company in the past years. As discussed in Section 2, various black-box test design approaches have been in use since the company was founded in 2010, e.g., category-partition testing and boundary-value testing. However, since such techniques can be interpreted and applied in different ways by different test engineers, the automated test suites were designed in different ways, and we have been observing imperfections in terms of test-case design (e.g., test duplications, and missing certain test cases or parts).

Based on our assessments (both qualitative and quantitate) during the project, since January 2019 up to this writing (Fall 2020), the key benefits of MBT have been as follows, which we discuss in detail next:

- Increased test effectiveness in detection of real faults
- Improved test-case design practices, due to MBT
- Ability to systematically assess requirements coverage by using MBT
- Intangible but important benefits

### 6.1.1 Increased test effectiveness in detection of real faults

While the SUT reported in this paper (Testinium, testinium.com) is a major product of the company with a few hundred clients, and that has been thoroughly tested throughout the years, we were still able to detect several major and minor issues in the SUT, via MBT. This was especially the case in the context of "regression" MBT, i.e., running the MBT suite every time that the SUT changes. In the last few months of the project (in 2020), there were a few updates to the SUT, and thus we re-executed the large MBT suite with no additional cost.

Since we have embedded the MBT execution in the company's Continuous Integration (CI) pipeline, in each revision of the SUT, the MBT test suite automatically runs at night, and it has detected 11 defects in the SUT in the duration of four months (June-September 2020, inclusive), e.g., system asking for login again after a long test execution, and certain test reports which were supposed to be shown by Testinium, not being displayed. The first above defect showed the value of endurance testing which was possible using MBT (each execution took more than 6 hours). Endurance testing (also known as soak testing) is a type of non-functional software testing. Endurance testing involves testing a system over a significant period of time, to discover how the system behaves under sustained use.

### 6.1.2 Improved test-case design practices, due to MBT

Test-case design before MBT was mostly ad-hoc. Although black-box test design approaches were slightly in use, as per inspections of the first author (an expert in software testing), many deficiencies were identified, e.g., duplications among different test suites (could lead to test integrity problems, etc.), and many missing test paths and test cases.

We wanted to compare objectively the improvement test-case design practices, due to MBT. Before introducing MBT, the SUT (Testinium) was tested automatically using a BDD test framework named Gauge (gauge.org), in which test scrips were written in natural language (mostly Turkish, but English would also be possible), and then Selenium code was developed to map those test scripts to Java commands to exercise the SUT. We have reported details of that other project in another recent paper [11]. For the interested reader, we show an example of a Gauge test-script for the same SUT (Testinium) in Table 4.

**Table 4-An example Gauge test-script for the Testinium as the SUT**

| | | | |
|---|---|---|---|
| Level of abstraction: High | 1 test "scenarios". Calling 2 test "concepts" | Tags:LoginPage_InputControls<br>* In the Login page, do the controls for the Email field<br>* In the Login page, do the controls for the Password field | Call |
| | 2 test "concepts". 3+4 test "steps" | # In the Login page, do the controls for the Email field<br>* Go to "http://www.testinium.com" address<br>* Check if the "tbEmailOfLogin" element is visible<br>* Login using "testinium" and "passwd**" | Call |
| | | # Login using <email> and <password><br>* Inside the field "tbEmailOfLogin", write <email><br>* Inside the field "tbPasswordOfLogin", write <password><br>* Check if the "btnSignInOfLogin" element exists<br>* "btnSignInOfLogin" elementine tıkla | Call the SUT |
| Level of abstraction: Low | A step "implementation" | @Step("Inside the field <key>, write <text>")<br>public void sendKey(String key, String text){<br>ElementInfo elementInfo = Stor<br>Helper.INSTANCE.findElementInfoByKey(key);<br><br>sendKeyBy(ElementHelper.getElementInfoToBy(elementInfo),<br>elementInfo.getIndex(), text);<br>} | |

We wanted to compare the two test-case design approaches: (1) **Before**: manual test design loosely following black-box testing approaches and test scripting / execution using Gauge, in which our test engineers loosely followed the black-box test design approaches, and (2) **After**: MBT (both test-case design and execution), as presented in this paper. Our metrics of comparison were: (1) number of real faults detected by each test suite, (2) the number of test steps generated in each approach, and (3) the time resource spent in each method to develop the test suites. The measured data for comparison are shown in Table 5. The values for the number of real faults detected have been gathered in the same settings, i.e., executing both test suites in a period of four months of the changed versions of the SUT (6 different versions).

As we can see in Table 5, manual test design using Gauge took about 4 engineer-months: one of the authors were developing the test suites. The MBT test suite development took 2 engineer-months. We can see that with less overall time resources, devoted to develop the test suites, MBT has provided more real-fault detection effectiveness. Also, in terms of test steps (comprehensiveness), MBT has enabled us to generate more than 10 fold more test steps (38,460 versus 2,658).

…

**Table 5-Comparing the two test-case design approaches**

| Test-design approach | Number of real faults detected | Number of test steps generated | Time resource spent to develop the test suites |
|---|---|---|---|
| **Manual test design loosely following black-box testing, and test scripting / execution using Gauge** | 3 | 2,658 | 4 engineer-months |
| **MBT (both test-case design and execution)** | 11 | 38,460 | 2 engineer-months |

As an example of what the MBT test suites and steps look like, we show in Figure 9 a partial snapshot from the output test log, an Excel file automatically generated by the extended reporting engine that we have added to the MBT tool (GraphWalker). We can see the start and the end of the very long sequence of 38,460 test steps in this log. As we can see, MBT test execution has taken more than 9 and half hours: starting on 22.34 PM and ending on 8:09 AM the next morning.

In Figure 9, the "model visit sequence number" column shows the sequence of visiting a given test model. As a reminder, the entire MBT suite of the SUT has been broken into multiple models (18 test models), as discussed in Section 5.3. The "Entity visit sequence number" column shows the sequence of visiting a given model entity (node or an edge of the state diagram). For example, some nodes and edges towards the end of the test log in Figure 9 have been visited more than 500 or 600 times.

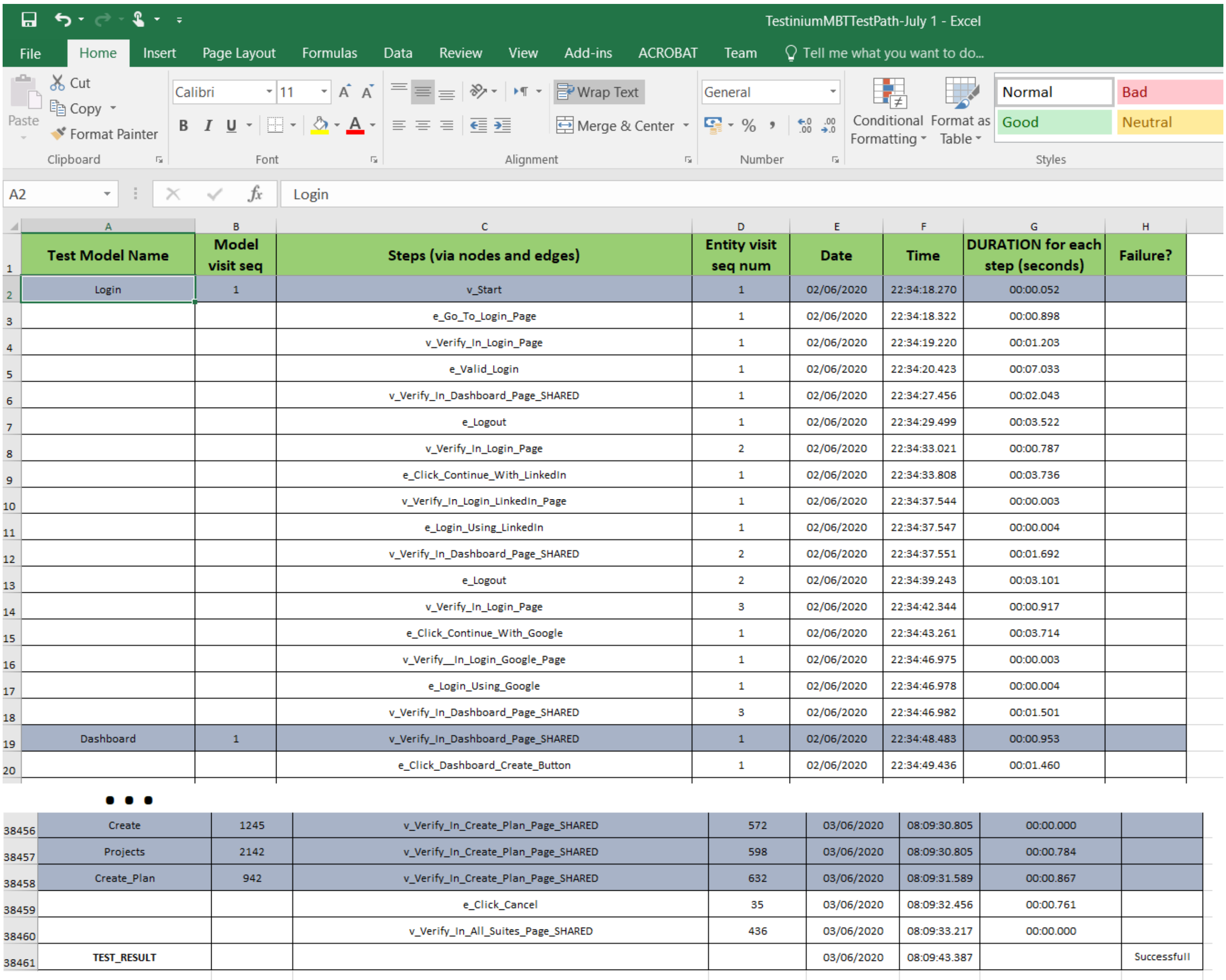

| | Test Model Name | Model visit seq | Steps (via nodes and edges) | Entity visit seq num | Date | Time | DURATION for each step (seconds) | Failure? |
|---|---|---|---|---|---|---|---|---|
| 2 | Login | 1 | v_Start | 1 | 02/06/2020 | 22:34:18.270 | 00:00.052 | |
| 3 | | | e_Go_To_Login_Page | 1 | 02/06/2020 | 22:34:18.322 | 00:00.898 | |
| 4 | | | v_Verify_In_Login_Page | 1 | 02/06/2020 | 22:34:19.220 | 00:01.203 | |
| 5 | | | e_Valid_Login | 1 | 02/06/2020 | 22:34:20.423 | 00:07.033 | |
| 6 | | | v_Verify_In_Dashboard_Page_SHARED | 1 | 02/06/2020 | 22:34:27.456 | 00:02.043 | |
| 7 | | | e_Logout | 1 | 02/06/2020 | 22:34:29.499 | 00:03.522 | |
| 8 | | | v_Verify_In_Login_Page | 2 | 02/06/2020 | 22:34:33.021 | 00:00.787 | |
| 9 | | | e_Click_Continue_With_LinkedIn | 1 | 02/06/2020 | 22:34:33.808 | 00:03.736 | |
| 10 | | | v_Verify_In_Login_LinkedIn_Page | 1 | 02/06/2020 | 22:34:37.544 | 00:00.003 | |
| 11 | | | e_Login_Using_LinkedIn | 1 | 02/06/2020 | 22:34:37.547 | 00:00.004 | |
| 12 | | | v_Verify_In_Dashboard_Page_SHARED | 2 | 02/06/2020 | 22:34:37.551 | 00:01.692 | |
| 13 | | | e_Logout | 2 | 02/06/2020 | 22:34:39.243 | 00:03.101 | |
| 14 | | | v_Verify_In_Login_Page | 3 | 02/06/2020 | 22:34:42.344 | 00:00.917 | |
| 15 | | | e_Click_Continue_With_Google | 1 | 02/06/2020 | 22:34:43.261 | 00:03.714 | |
| 16 | | | v_Verify__In_Login_Google_Page | 1 | 02/06/2020 | 22:34:46.975 | 00:00.003 | |
| 17 | | | e_Login_Using_Google | 1 | 02/06/2020 | 22:34:46.978 | 00:00.004 | |
| 18 | | | v_Verify_In_Dashboard_Page_SHARED | 3 | 02/06/2020 | 22:34:46.982 | 00:01.501 | |
| 19 | Dashboard | 1 | v_Verify_In_Dashboard_Page_SHARED | 1 | 02/06/2020 | 22:34:48.483 | 00:00.953 | |
| 20 | | | e_Click_Dashboard_Create_Button | 1 | 02/06/2020 | 22:34:49.436 | 00:01.460 | |
| | • • • | | | | | | | |
| 38456 | Create | 1245 | v_Verify_In_Create_Plan_Page_SHARED | 572 | 03/06/2020 | 08:09:30.805 | 00:00.000 | |
| 38457 | Projects | 2142 | v_Verify_In_Create_Plan_Page_SHARED | 598 | 03/06/2020 | 08:09:30.805 | 00:00.784 | |
| 38458 | Create_Plan | 942 | v_Verify_In_Create_Plan_Page_SHARED | 632 | 03/06/2020 | 08:09:31.589 | 00:00.867 | |
| 38459 | | | e_Click_Cancel | 35 | 03/06/2020 | 08:09:32.456 | 00:00.761 | |
| 38460 | | | v_Verify_In_All_Suites_Page_SHARED | 436 | 03/06/2020 | 08:09:33.217 | 00:00.000 | |
| 38461 | TEST_RESULT | | | | 03/06/2020 | 08:09:43.387 | | Successfull |

**Figure 9- Partial snapshot from the output test log, generated by the extended reporting engine that we have added to the MBT tool**

Thus, it is clear that MBT has led to major improvements in test-case design practices. Furthermore, we were able to gather several expert opinions from senior testers who confirmed that test-case design practices have indeed been improved, thanks to MBT. As a qualitative feedback, one test manager mentioned that: "*The new MBT approach is very promising and we are excited to use it more widely in as many test projects across the company as possible. MBT has increased effectiveness and efficiency of our test design activities and will continue to do so in future.*"

### 6.1.3 Ability to systematically assess requirements coverage by using MBT

As discussed in Section 5.5, one of the work packages of the R&D project (testomatproject.eu) was to measure requirement coverage in design and also execution of test suites and also incorporating test-requirement traceability. We used the MBT tool's simple but effective feature for requirements coverage and traceability (as shown in Figure 6). That pragmatic and lean feature helped us and our test engineers to design the MBT models while tracing certain elements of test models (nodes and edges) back to requirements. Upon finishing execution of MBT test suites, since we had chosen edge-coverage of models = 100% as the test termination criteria, the tool would also provide 100% requirement coverage as the outcome. That test-requirement traceability approach was welcomed by our test engineers and managers, and indeed was seen as a benefit and (positive) impact of MBT.

### 6.1.4 Intangible but important benefits

According to the informal feedback of the engineers involved in this project, MBT made the work of test engineers more "interesting", and more organized. Many in the company have told us that, thanks to MBT models, they can now see the "big picture" of test-case design much more easily with having the test models in front of them, and the model being directly executable. Attention to intangible benefits of any Software Engineering (SE) approach in practice is widely discussed (bit.ly/Intangible_BenefitsInSE) [26, 60]. While such benefits are difficult to measure, they should be considered when assessing usefulness of any SE approach in practice. We were curious to know what made MBT and MBT test development more "interesting" to our engineers and when, we asked them, we were told that development of MBT models looks like an innovative and joyful design development task (see the model examples in Figure 4), and engineers would learn more about test automation and the SUT in the process. It was clear to us and our test engineers that such intangible benefits were indeed very important.

**Challenge**: Systematic (quantitative) assessment and comparison of improvements in test-case design practices in our industrial setting has not been trivial, due to issues such as measurements in work practices seen as "extra work" and also to sensitivity of such measurements (outcome of measurements could harm practitioners' prestige and position), as reported in other studies too [61]. In other words, from an academic standpoint, formally quantify the benefits of MBT in industry is a challenge, as controlled experiments are often not feasible in this context (not easy from an industrial standpoint to "justify" the time and effort to do such experiments). We are exploring ways of doing such comparisons in a quantitative but still viable ways.

**Lesson learned**: MBT provides various intangible benefits which are as important as if not more as its tangible (measurable) benefits. It is important to recognize, analyze and discuss them in test teams. As Albert Einstein said: "*Not everything that counts can be counted*" [62].

We were also keen and careful of the cost-effectiveness of MBT throughout the MBT project so far. We are also aware of several other studies which have touched on this issue, e.g., an experience report [13] by a researcher who had worked in industry for several years and had used MBT, mentioned three points in this regard: (1) "*it's risky to choose an MBT approach without having a clear view about its complexity, cost, effort, and skill required to create [develop] the necessary models*" [8]; (2) "*it is important to always state where the models [to be used in model-based testing] come from: are they artificial or did they already exist before the experiments*" and (3) "*one has to argue and evaluate if the time and effort in developing and maintaining such models for a given system does pay off in the end*". For this purpose, we had group and individual discussions with test engineers in our MBT project in several iterations. All the involved team members agreed that costs invested into MBT have been well worth it, and everyone is eager to see a wider adoption of MBT in many client projects in the company.

**Lesson learned**: The action-research project (and the case study) of introducing MBT at Testinium A.Ş. has been a success so far, as all involved engineers and managers found it useful, and want to continue to use it, and want to apply it to other products as well.

### 6.2 Challenges and open questions observed so far (ARQ3)

During the entire project, we have gathered the challenges and the open questions that we have observed so far, as reported throughout the paper in the previous section. We provide some further discussions on each of them.

**Challenges:**

- While there are some studies which have proposed design patterns for UML models in general [63], we observed a general shortage of knowledge and resources on best practices and design patterns for designing MBT models (as discussed in Section 5.3). We thus recommend more research and investigations on this very important topic by researchers and practitioners in future. Our approach for quality assurance of MBT models was regular peer review and inspection by experience team members, but there is need for design patterns for this purpose to help test engineers new to MBT design to design high-quality MBT models.
- Systematic quantitative assessment and comparison of improvements in test-case design practices in our industrial setting has not been trivial (Section 6.1), due to issues such as measurements in work practices seen as "extra work" from industrial side and also due to sensitivity of such measurements (outcome of measurements could harm practitioners' prestige and position), as reported in other studies too [61]. In other words, from an academic standpoint, formally quantifying the benefits of MBT in industry is a challenge, as controlled experiments are often not feasible in this context (not easy form industrial standpoint to "justify" the time and effort to do such experiments). We are still exploring ways of doing such comparisons in a quantitative but still viable cost-effective ways, so that we could get the buy-in of the practitioners.

**Open questions:**

- Any given MBT tool and approach (including the one that we selected: GraphWalker), has various parameters and configurations to be set, such as the choice of graph traversal algorithms to generate test paths (Section 5.7). An important open question is which sets of parameters are the best, or would provide better test outcomes, coverage, execution time, fault detection effectiveness, etc. This raises the need for empirical studies on the choice of those parameters and configurations and also possibly some new Search-based Software Engineering (SBSE) [58] approaches to select the best options.

## 7 Discussion: Lessons learned, limitations and take-away messages

We discuss the lessons learned, limitations and take-away messages in this section.

### 7.1 Increase in maturity and capability of test automation using MBT

In software engineering and software testing, maturity and capability of teams and organizations are widely discussed issues. Usually, maturity relates to how well a team or a organization performs all processes of a software engineering topic, e.g., testing [64]. Capability, on the other hand, relates to and is the level of a team or an organization's ability and improvement achievement in specific technical area, e.g., MBT testing.

In our context, test automation maturity was the high-level "umbrella" notion, under which we wanted to improve various automation "capabilities". Higher capabilities in MBT has been one of the approaches to increase the company's overall test automation maturity. As we discussed in the paper's abstract and Section 1, in the planning phase of this project, we were keen to improve both maturity and capability of test automation using MBT in the subject company.

There are systematic maturity and capability improvement models such as Test Maturity Model integration (TMMi) [65-67]. We also have some past experience in applying TMMi in other industrial contexts, e.g., see the case study section of [68].

For the current paper and the current MBT project that we have conducted in the specific industrial context (Testinium A.Ş.), we did not decide to use such systematic maturity improvement models in this first phase of the project, but instead we decided to focus on the "technical" aspects of the MBT (as discussed in Section 5), and also evaluating the benefits and challenges of the MBT approach (Section 6). Certainly, as a by-product, as both the tangible and intangible benefits of MBT showed (Section 6.1), it is certain that maturity and capability of test automation in the subject company have improved using MBT. We have plans to utilize one or more of the systematic maturity and capability improvement models, from the literature, in this industrial context in future, and to assess whether they would provide extra benefits and improvement recommendations.

### 7.2 Lessons learned and Take-away messages

We summarize below the main lessons that we have learned so far in our MBT project, and also the practical advice based on our described experience. Let us clarify that these lessons learned shall not be interpreted as facts or general rules, but they are only several main lessons learned, experience and beliefs from our particular project and context.

**Lesson learned 1:** We empirically observed that choosing the "right" MBT tool from amongst the very large pool of available MBT tools is not trivial (Section 5.2). We found that, as also reported in many other resources, selecting the "right" tool for the "right" purpose in MBT is a key to success. We found the guidelines of a Grey-Literature Review (GLR) [2] in this topic useful as they helped us choose the right tool. When introducing MBT to a company for the first time, a lightweight MBT tool/approach is advisable, especially when there exist success stories from other practitioners that have successfully used a given MBT tool in other industrial contexts (companies).

**Lesson learned 2:** Even if the MBT models may be developed automatically by reverse-engineering them from the web SUT, we found that manual development of MBT models by test engineers provided various "side" benefits, e.g., valuable learning experience, increasing motivations and interest of test engineers in test automation (Section 5.3). There were indeed some of the intangible benefits of MBT (Section 6.1).

**Lesson learned 3:** The modeling semantic of the chosen MBT tool provided (in a sense, "enforced") a suitable "separation of concerns" (SoC) (design pattern) in a way to make the test code modular and helped test engineers clearly know what to develop for each Java method. This shows the importance of choosing the right MBT approach and tool (Section 5.2).

**Lesson learned 4:** Including the MBT execution in the Continuous Integration (CI) pipeline was shown to be a good practice (Section 5.7), as it would execute automatically regularly (every night) and report the results.

**Lesson learned 5:** While there are numerous coverage tools for conventional test automation, e.g., for xUnit frameworks, to our surprise, there were no off-the-shelf readily-applicable coverage tools to work in conjunction with MBT for web applications, which would derive and show both front-end (client-side) JavaScript (JS) and also back-end (server-side) coverage values, "live", as a given MBT test suite is running. Thus, practitioners should be aware of this, if they plan to measure coverage in conjunction with MBT. The *MBTCover* tool (Section 5.8), that we have developed, works with the chosen MBT tool (GraphWalker), and thus it can be helpful.

**Lesson learned 6:** The action-research project of introducing MBT at Testinium A.Ş. has been successful so far, as all involved engineers and managers found it useful, and want to continue to use it, and want to apply it to other test projects as well (Section 6.1).

Furthermore, as discussed in Section 1, several motivators for this experience reports were the following phrases from the literature, and for each of them we provide how this experience report relates or contributes to:

- "*…a serious lack in evidence*" in MBT [25]: We have first-hand observed the evidence that MBT can work well in practice if planned and conducted carefully, as we have done in our project. We thus believe this paper contributes to the body of empirical evidence in industrial application of MBT by sharing our industry-academia project on applying MBT in practice.
- "*most developers [still] don't view MBT as a mainstream [testing] approach*" [8] (a paper in 2008): Although as per review of the industry and grey literature of MBT, industrial adoption of MBT still seems limited as of 2020, sharing positive evidence of MBT in practice can help change the situation in the position direction, one success story at a time, and gradually bring MBT to "mainstream". In addition to sharing our experience in academic papers like this, we actively disseminate our experience and findings to practitioners via industrial talks, e.g., [69]. We have been able to "influence", in a good way, at least five test engineers in our company to start using MBT in an active mode to develop test suites.
- "*Developers must obviously take care to select an MBT approach that matches their project's specific needs*" [8]: By carefully considering our industrial context and project needs (Section 2), we selected the right test automation approach and tool (Section 5.2), and we invite all practitioners to use our example approach in their projects.
- "*it is important to always state where the models [to be used in MBT] come from: are they artificial or did they already exist before the experiments*" and "*one has to argue and evaluate if the time and effort in developing and maintaining such models for a given system does pay off in the end*" [13]: For this issue, it is true that we developed the MBT models from scratch since they did not exist from before in our project, but with a pragmatic / lean MBT model formalism and good usability, the MBT tool has minimized our MBT model development efforts (Section 5.3),and all team members believe that efforts put into MBT model development are well worth it, especially since manual development of MBT models by test engineers provided various "side" benefits, e.g., valuable learning experience, increasing motivations and interest of test engineers in test automation. We believe this is in quite a contrast to many MBT approaches, proposed in academia which are often not cost-effective in industry, e.g., the PhD work of the first author [70]. We discussed in some recent works [19, 71] why such "heavy-weight" MBT approaches are hard to be applied in practice (industry).

### 7.3 Limitations

The action-research project of introducing MBT at Testinium A.Ş. has been successful so far. However, no project in industry or academia can be "perfect". We have been aware of the limitations of our project so far, and in fact, we are working on them. These limitations are:

**Limitation 1:** As the first paper of the MBT project, the current paper is an "experience report" based on action-research [20-23], in which we have synthesized and presented the project, and empirical benefits of MBT in practice. To increase the rigor of our assessment, we plan to conduct controlled experiments and rigorous case studies in our ongoing R&D project in future.

**Limitation 2:** Choice of graph traversal algorithms in the GraphWalker tool to generate test paths (Section 5.7): To keep the complexity of our work in a manageable level so far, we have configured the MBT runs to use the "Random" option. We plan to conduct in-depth studies by using the other graph traversal algorithms (choice of generators).

**Limitation 3:** In this paper, we presented our experience of MBT in the context of one single SUT (Testinium, Figure 3). Work has already started in the company to apply MBT on more large-scale SUTs, and we plan to share more findings from those efforts as they become available in future.

**Limitation 4:** By seeing the lack of proper MBT coverage tools, development of our MBT coverage tool (MBTCover), as discussed in Section 5.8, has been a good step in the project. While we have used that tool in an "exploratory" manner to assess coverage and improve MBT test suites as needed, we plan to conduct more systematic studies on MBT coverage.

**Limitation 5:** We should also highlight that we applied MBT in one particular setting and thus every practitioner should carefully review the literature (at the end of this paper), and do a proper planning to prevent disappointment with MBT. Based on our experience, we can recommend the use of lightweight tools such as GraphWalker as a first step to introduce MBT in companies dealing with *web* and *enterprise applications*. As there are a very few reported research and success stories in this domain, this experience report provides a valuable contribution to inspire practitioners to try out MBT on their projects.

**Limitation 6:** The opinion bias of the authors and the interviewed employees on the results can be a factor and threats to validity.

## 8 Conclusions and ongoing/future works

We are glad to share that MBT has fully fulfilled the expectations in our industrial context. We believe that a good strategy and a pragmatic approach has enabled us to achieve this. Our experience also confirmed that following lesson learned in [25]: "*Because of the complexity of MBT in comparison to existing testing techniques, any actions that aim at spreading MBT inside a company have to be taken in small steps*".

Based on the feedback from the engineers that worked in this project, our MBT project has showed us several important ongoing work directions, e.g.: (1) we are in the process of improving the testing tool to incorporate fault tolerance (when an assertion fails), (2) test visualization (showing the number of times each edge and node has been covered), (3) quantitative assessment of the benefits of MBT; and (4) assessing effectiveness of MBT in detection of injected faults (by mutation testing).

## Acknowledgements

This work was supported by the European ITEA3 program via the "TESTOMAT (The Next Level of Test Automation)" project with grant number 16032, by the Scientific and Technological Research Council of Turkey (TÜBİTAK) with grant number 9180076, and also by the Research Council of Norway with grant number 274385.